\documentclass{IEEEtran}

\ifCLASSINFOpdf
   \usepackage[pdftex]{graphicx}
\else
   \usepackage[dvips]{graphicx}
\fi

\usepackage[cmex10]{amsmath}
\usepackage{bm}

\usepackage{algorithmic}
\usepackage{algorithm}
\usepackage{cancel}
\usepackage{cleveref}
\usepackage[table]{xcolor}

\usepackage{caption} 
\usepackage{subcaption}

\newfloat{model}{thp}{lop}
\floatname{model}{Model}

\begin{document}

\newcommand{\com}[1]{\begin{code}{#1}\end{code}\nonumber}

\newcommand{\bus}{\mathcal{B}}
\newcommand{\busid}{i}
\newcommand{\busidtwo}{j}

\newcommand{\gen}{\mathcal{G}}
\newcommand{\genid}{g}

\newcommand{\branch}{\mathcal{L}}

\newcommand{\demand}{\mathcal{D}}
\newcommand{\demandid}{d}

\newcommand \nr[1] {{\color{red}#1}}
\newcommand{\rev}[1]{{\color{red}#1}}
\newcommand{\note}[1]{{\color{blue}#1}}

\title{Recursive Restoration Refinement: \\ A Fast Heuristic for Near-Optimal Restoration Prioritization in Power Systems}

\author{\IEEEauthorblockN{Noah Rhodes\IEEEauthorrefmark{1}\IEEEauthorrefmark{2},Carleton Coffrin\IEEEauthorrefmark{2}, and Line Roald\IEEEauthorrefmark{1}}
\IEEEauthorblockA{\IEEEauthorrefmark{1} University of Wisconsin-Madison, Madison, Wisconsin, USA}
\IEEEauthorblockA{\IEEEauthorrefmark{2} Los Alamos National Laboratory, Los Alamos, New Mexico, USA}
\thanks{\noindent This work was supported by funding from the U.S. Department of Energy’s (DOE) Office of Electricity as part of the CleanStart-DERMS project of the Grid Modernization Laboratory Consortium and by the DOE Office of Science, Office of Advanced Scientific Computing Research under Contract DE-AC02-06CH11357. LA-UR-21-29872}}

\maketitle

\begin{abstract}
The prioritization of restoration actions after large power system outages plays a key role in how quickly power can be restored.
It has been shown that fast and intuitive heuristics for restoration prioritization most often result in low-quality restoration plans.
Meanwhile, mathematical optimization tools that find high-quality restoration plans are too slow to be applied to restoration planning problems of practical interest.
This work makes a significant step in closing this quality vs compute time gap by proposing the \emph{Recursive Restoration Refinement} heuristic for power system restoration.
This heuristic is shown to produce near-optimal restoration plans up to 1,000 times faster than other state-of-the-art solution methods on a range of test cases with up to 500 buses and 700 damaged components.
The potential impact of this new heuristic is demonstrated by a preliminary analysis of the key features of high-quality restoration plans. 
The recursive restoration refinement algorithm and other methods explored in this work have been made available as part of the open-source software package, PowerModelsRestoration, to support ongoing research in power restoration algorithms. 
\end{abstract}

\begin{IEEEkeywords}
Power system restoration algorithms, component restoration prioritization, 
mixed-integer optimization
\end{IEEEkeywords}

\vspace{-0.5cm}

\section{Introduction}

The electric grid is susceptible to outside threats ranging from natural disasters such as hurricanes and wildfires to targeted attacks. Quickly restoring the transmission network after a large system-wide failure is important as we increasingly rely on electricity in all aspects of life.
To achieve this restoration as quickly as possible, utilities need restoration planning tools that can assist in the design and execute high-quality restoration plans \cite{6516097}.

Restoration planning is a broad topic that considers a wide range of operating situations and time scales. As a result, there are a variety of specific task and operational problems that are studied as part of it \cite{LINDENMEYER2001219}. 
These tasks and problems include topics like managing black-start capable generators \cite{JIANG2017127} \cite{5648472}, cold-load pick up \cite{7286872}, transient stability \cite{Hijazi_Mak_VanHentenryck_2015}, dispatching restoration crews \cite{van2011vehicle} \cite{8640043}, coordination of parallel restoration activities \cite{7932458} \cite{en11051316}, and considering varying duration of restoration actions \cite{YAN2020105612} \cite{6246657}.  
Other work considers the potential for creating microgrids to serve critical load \cite{7513408} or using restoration planning to inform grid hardening \cite{8359327}. Many of these problems must be addressed for a utility to make restoration decisions, and developed into decision support tools \cite{kirschen1991guiding} \cite{adibi1987taskforce}.  These problems can be addressed in stages, as restoration milestones are achieved \cite{5355208}.
Power system restoration can include \emph{physical repair} of components, e.g., following a hurricane, and/or efforts to \emph{reenergize and reconnect} the system after a large disruption, e.g., a system-wide blackout. Here, we focus on the former type of problem where the grid has sustained physical damage.
Specifically, we focus on the Component Repair Priority (CRP) problem, which is the task of determining where to focus repair efforts over time to maximize the amount of power that can be delivered. Prioritizing components for repair is important, as there are typically only limited resources (i.e., crews and equipment) available to perform the repairs and we therefore need to decide on which components to repair first.
This CRP problem is of particular interest as it occurs in a variety of restoration planning activities and also presents a daunting computational challenge. 

In previous related work \cite{rhodes2021powermodelsrestoration}, we presented a software implementation of the Restoration Ordering Problem (ROP), an optimization problem to solve the CRP problem,  which allows for testing of solution times and solution quality with AC, SOC, and DC power flow formulations. This work also discussed a preprocessing heuristic from \cite{VanHentenryck2015} to reduce the number of components considered for restoration prioritization. However, we observed that the ROP problem was very challenging to solve even with DC power flow and a moderate number of damaged items. Therefore, this paper focuses on scalable heuristic algorithms that obtain near-optimal solutions to large-scale CRP in a short amount of time. 
We consider only line damage and use a DC power flow model
\cite{4956966}, as even this simplified version of CRP presents a significant computational challenge.

The core contribution of this work is a new heuristic algorithm for solving the CRP problem called \emph{Recursive Restoration Refinement (RRR)}.
In particular, it is shown that the RRR method can produce near-optimal solutions for the CRP tasks up to 1,000 times faster than the current state-of-the-art.
This result is achieved by careful comparisons of RRR to established methods, including prioritizing repairs based on component utilization \cite{coffrin2012last}, obtaining the optimal solution by solving the \emph{Restoration Ordering Problem} (ROP) \cite{VanHentenryck2015}, and using a state-of-the-art meta-heuristic  \emph{Randomized Adaptive Decomposition } (RAD) \cite{coffrin2012last}.  These methods are compared on 6 grids ranging from 24 to 500 buses, each with 10 damage scenarios, and solved with a 5-minute and 10-hour time limit.
Overall, the RRR algorithm significantly improves the scalability and quality of solutions for the CRP problems. This provides a new avenue for designing and understanding complex restoration plans, as illustrated by a preliminary analysis of high-quality restoration plans of a 240 bus system with more than 200 damaged components.

The paper reviews the CRP problem and previous approaches to solving it in Section \ref{sec:CRP}.  We then introduce the proposed \emph{RRR} algorithm in Section \ref{sec:RRR}, and some post-processing details in Section \ref{sec:processing}. Section \ref{sec:results} presents numerical evaluations of the algorithm, and  Section \ref{sec:conclusion} concludes.

\section{The Component Repair Priority Problem}
\label{sec:CRP}

This section first introduces the CRP problem, before reviewing three existing approaches for solving it, the Utilization Heuristic, the Restoration Ordering Problem, and the Randomized Adaptive Decomposition algorithm.

\subsection{Line Restoration Plans}
\label{sec:brp}

The CRPs of interest to this work consist of a power grid $\cal N$ and a set of physically damaged lines $\bcancel{\cal L}$.
The computational task is to find a line restoration sequence of the lines in $\bcancel{\cal L}$ that maximizes the amount of power served over time.

\subsubsection{Restoration Plan Definition}
Noting that any line in the network $\cal N$ can be uniquely identified by a pair of buses $(i,j)$, a restoration plan $\mathcal{R}$ is defined as a vector of restoration periods, $\mathcal{R} = \left[ \mathcal{R}_1, ..., \mathcal{R}_k, ..., \mathcal{R}_N \right]$, where each period $k$ contains a set of lines that are restored in that period, e.g., $\mathcal{R}_k = \langle (i,j), (n,m) \rangle $.
For example, a restoration plan with 5 restored lines and $N=5$ could be ${\cal R} = \left [ \left \langle (1,2)  \right \rangle, \left \langle \right \rangle, \left \langle (1,4), (2,3) \right \rangle, \left \langle (4,5)\right \rangle, \left \langle (1,5)\right \rangle \right ]$.
Note that this definition supports a different number of restored components in each time step, i.e., energization of zero, one or more components in each period $k$, as is necessary or advantageous to overall power delivery.

\subsubsection{Power Delivery of Restoration Plan}
Given a restoration sequence $\mathcal{R}$, we can compute the maximum power delivered in each restoration period by solving a continuous, multi-period optimal power flow problem that aims at maximizing total power delivery. This problem is defined  Model \ref{model:rip} and we refer to is as the \emph{Restoration Implementation Problem} (RIP). 

The objective function of RIP \eqref{eq:rip_objective} maximizes the total power delivered. This is defined as the product of $x^D_{dk}$, denoting the fraction of load that can be served at node $d$, by the total power demand of that node ${\boldsymbol{P^D}_d}$ and the duration of the time step $\mathbf{\Delta}_k$, summed over all loads $d \in \mathcal{D}$ and all repair periods $k \in 1,...,N$.  ${\boldsymbol{P^D}_d}$ is constant and assumed to be the maximal load, but can also be indexed by $k$ if a time-varying load forecast is available.
For each restoration period, the power balance at each node is enforced by constraint \eqref{eq:rip_power_balance}, where the power from all connected generators $P^G_{gk}$, lines $P^L_{ijk}$, and loads ${x_{dk}\boldsymbol{P^D}_d}$ must sum to 0 at each bus $i \in \mathcal{B}$ .  The line flows are expressed by the DC power flow formulation in \eqref{eq:rip_flow_limit} with $\theta_{ik}$ representing the voltage angle at bus $i$ in restoration period $k$ and $\boldsymbol{b}_{ij}$ represent the line susceptance. Eq. \eqref{eq:rip_thermal_limit} enforces the power flow limits $\overline{\boldsymbol{P^L}_{ij}}$.  These constraints are only included for lines that are energized in restoration period $k$. The set of energized lines denoted by $\mathcal{L}_k$ comprises both non-damaged lines and the lines restored in the previous periods $\mathcal{R}_1,\dots,\mathcal{R}_k$.  Finally, constraint \eqref{eq:rip_gen_limits} enforces the upper and lower bounds $\overline{\boldsymbol{P^G}_g}$ and $0$ on the generator power $P^G_{\genid k}$ for all generators $g \in \mathcal{G}$.
The lower bound on generation is 0 to ensure there is a feasible generator dispatch for any amount of load in any period of the restoration \cite{coffrin2019}.  We leave the modeling of non-zero generator lower bounds and generator commitment to future work.

We assume the physical repairs require time to be conducted and the restorations occur on the order of hours, so we neglect stability concerns and consider a series of static power flow problems.  We also neglect temporal constraints related to generator ramping and standing phase angle constraints. It has been shown that standing phase angle constraints can be incorporated with a relatively small impact on the amount of power served during the restoration process,  less than 1.5\%
\cite{7038388}.
\begin{model}[t]
\caption{Restoration Implementation Problem (RIP)}
\label{model:rip}
\begin{subequations}
\vspace{-0.2cm}
\begin{flalign*}
& ~~ \mbox{\bf variables: ($\forall k \in 1,...,\boldsymbol{N}$)}  & \\
& ~~ P_{\genid k}^G \; \forall \genid \in \gen,  ~
            P_{ij k}^{L} \; \forall (i,j) \in \branch_k,   ~
            x^D_{d k} \; \forall \demandid \in \demand, ~
            \theta_{i k} \forall i \!\in\! \mathcal{B} &
\end{flalign*}
\begin{align}
& \mbox{\bf maximize:}  \quad \sum_{k \in 1,...,\boldsymbol{N}}  \sum_{\demandid \in {\demand}} x^D_{\demandid k}   \boldsymbol{P}^D_{\demandid} \boldsymbol{\Delta} _k
\label{eq:rip_objective}\\
&\mbox{\bf subject to ($\forall k \in 1,...,\boldsymbol{N}$):} \nonumber \\
& \sum_{g\in\mathcal{G}_i} P_{g k}^G +\!\!\! \sum_{(i,j)\in\mathcal{L}_{ik}}\!\!\!P_{ k}^L -\!\!\! \sum_{d\in\mathcal{D}_i} x^D_{d k} \boldsymbol{P}^D_d = 0 && \forall i \in \mathcal{B}
\label{eq:rip_power_balance}\\
& P_{ij k}^{L} = -\boldsymbol{b}_{ij} (\theta_{i k} - \theta_{j k}) && \forall (i,j) \in \mathcal{L}_k
\label{eq:rip_flow_limit} \\
&-\overline{\boldsymbol{P^L}_{ij}} \le  P_{ijk}^{L} \le \overline{\boldsymbol{P^L}_{ij}} && \forall (i,j) \in \mathcal{L}_k
\label{eq:rip_thermal_limit}\\
& 0 \le P_{\genid k}^G \le  \overline{\boldsymbol{P^G}_{\genid}} && \forall g \in \gen
\label{eq:rip_gen_limits}
\end{align}
\end{subequations}
\end{model}
We next review three approaches to solve the CRP problem and identify good restoration plans $\mathcal{R}$.

\subsection{Utilization Heuristic}
We first introduce the Utilization (UTIL) heuristic, a greedy heuristic that reflects the established industry practice of \emph{restoring the largest lines first} \cite{coffrin2012last}.
Algorithm \ref{alg:util} provides an implementation of the UTIL heuristic that takes a power grid $\cal N$ and a set of damaged lines $\bcancel{\cal L}$ as an input and returns a restoration plan $\cal R$ where all damaged lines are restored one-by-one in order of decreasing power capacity $\overline{\boldsymbol{P^L}_{ij}}$.

UTIL represents an intuitive
approach to the CRP, as
one expects that large capacity lines, which make up the backbone of the power grid, should be restored first to quickly increase the maximal capacity of the grid.
However, as previous work \cite{coffrin2012last} and the results in Section \ref{sec:results} highlight, UTIL tends to produce restoration plans that are far from optimal in terms of power delivery.
These observations indicate that accounting for the network connectivity and the power flow of the system is critical to ensure that the power line capacity is effectively utilized during the restoration process.

\begin{algorithm}[t]
    \caption{Utilization Heuristic (UTIL)}
        \begin{algorithmic}[1]
        \renewcommand{\algorithmicrequire}{\textbf{Input:}}
        \renewcommand{\algorithmicensure}{\textbf{Output:}}
        \REQUIRE $\mathcal{N}, \bcancel{\mathcal{L}}$ 
        \ENSURE  $\mathcal{R}$\\ 
        \STATE {$P \leftarrow [ (i, j, \overline{\boldsymbol{P^L}_{ij}}) \;\; \forall (i,j) \in \bcancel{\branch} ]$}
        \STATE $ S \leftarrow sortByLargestCapacity(P)$
        \STATE $\mathcal{R} \leftarrow [\mathcal{R}_k = \langle \rangle \;\; \forall k \in 1...|\bcancel{\cal L}|]$
        \STATE $k \leftarrow 1$
        \FOR {$(i,j,\boldsymbol{P^L}_{ij}) \in S$}
            \STATE $\mathcal{R}_k \leftarrow \langle (i,j) \rangle $
            \STATE $k \leftarrow k+1$
        \ENDFOR
        \RETURN $\mathcal{R}$
    \end{algorithmic}  \label{alg:util}
\end{algorithm}

\subsection{Restoration Ordering Problem} \label{sec:ROP}
We next consider the Restoration Ordering Problem (ROP) proposed in \cite{van2011vehicle}, which is a mathematical optimization model that can be solved with commercial mixed-integer programming solvers to find globally optimal solutions to the CRP.
At a high level, the ROP is a multi-period power flow problem where components are restored incrementally to maximize the amount of power that can be delivered to the loads overtime.

\begin{model}[t]
\caption{Restoration Ordering Problem (ROP)}
\label{model:rop}
\begin{subequations}
\vspace{-0.2cm}
\begin{flalign*}
&  \quad \mbox{\bf variables: ($\forall k \in 1,...,\boldsymbol{N}$)}  &\\
& \quad P_{\genid k}^G  \forall \genid\! \in\! \gen,  ~
            P_{ij k}^{L}  \forall (i,j) \!\in\! \branch,   ~
            z^L_{ijk}  \forall (i,j) \!\in\! \bcancel{\branch},  \\
& \quad
            x^D_{d k}  \forall \demandid \!\in\! \demand, ~
            \theta_{i k} \forall i \!\in\! \mathcal{B}  &
\end{flalign*}
\begin{align}
& \mbox{\bf maximize:}  ~ \text{\eqref{eq:rip_objective}} & \nonumber \\
&\mbox{\bf subject to ($\forall k \in 1,...,\boldsymbol{N}$):} \nonumber \\
& \sum_{ij \in \bcancel{\branch}}z^L_{{ij} k} \le \boldsymbol{R}_k
\label{eq:repair_limit} \\
& z^L_{ijk}\! \le \!z^L_{ij(k+1)} \;\; &&  \mkern-86mu \forall (i,j) \in \bcancel{\branch}, ~ \text{for } k\neq {\boldsymbol{N}}
\label{eq:repair_active}\\
& {z^L_{ijk}}\! = 1 \;\; && \mkern-86mu \forall (i,j) \in\bcancel{\branch}, ~ \text{for } k = {\boldsymbol{N}}
\label{eq:repair_all} \\
& P_{ij k}^{L} = -\boldsymbol{b}_{ij} (\theta_{i k} - \theta_{j k}) &&  \mkern-86mu \forall (i,j) \in \mathcal{L} \setminus \bcancel{\mathcal{L}} 
\label{eq:flow_limit_u} \\
&-\overline{\boldsymbol{P^L}_{ij}} \le  P_{ijk}^{L} \le \overline{\boldsymbol{P^L}_{ij}} && \mkern-86mu \forall (i,j) \in \mathcal{L} \setminus \bcancel{\mathcal{L}} 
\label{eq:thermal_limit_u}\\
& P_{ij k}^{L} \le -\boldsymbol{b}_{ij} (\theta_{i k} - \theta_{j k}) + \overline{\boldsymbol{\theta^{\Delta}}_{ij}} (1-z^L_{ij k}) && \forall (i,j) \in \bcancel{\mathcal{L}}
\label{eq:flow_limit1_d} \\
& P_{ij k}^{L} \ge -\boldsymbol{b}_{ij} (\theta_{i k} - \theta_{j k}) + \underline{\boldsymbol{\theta^{\Delta}}_{ij}} (1-z^L_{ij k}) && \forall (i,j) \in \bcancel{\mathcal{L}}
\label{eq:flow_limit2_d}\\
&-\overline{\boldsymbol{P^L}_{ij}}z^L_{ijk} \le  P_{ijk}^{L} \le \overline{\boldsymbol{P^L}_{ij}}z^L_{ijk} && \forall (i,j) \in \bcancel{\mathcal{L}}
\label{eq:thermal_limit_d}\\
& \sum_{g\in\mathcal{G}_i} P_{g k}^G +\!\!\! \sum_{(i,j)\in\mathcal{L}_i} P_{ k}^L -\!\!\! \sum_{d\in\mathcal{D}_i} x^D_{d k} \boldsymbol{P}^D_d = 0 && \forall i \in \mathcal{B}
\label{eq:power_balance} \\
& ~ \text{Generators Limits: \eqref{eq:rip_gen_limits}} \nonumber
\end{align}
\end{subequations}
\end{model}

The Restoration Ordering Problem (ROP) considered in this work is shown in Model \ref{model:rop}.
The formulation consists of $\boldsymbol{N}$ restoration periods. 
The primary decision variable is the status of each damaged line in $\bcancel{\cal L}$, which is represented by a binary variable $z^L_{ijk}$, indexed by the connecting buses $i,j$ and the restoration period $k \in 1,...,\boldsymbol{N}$. If $z^L_{ijk}=1$ (or $z^L_{ijk}=0$), the line is restored (or still damaged).
The objective function \eqref{eq:rip_objective} is the same as in the RIP problem and maximizes the total energy served across all time periods.

Eq. \eqref{eq:repair_limit}-\eqref{eq:repair_all} constrain how lines can be restored.
Eq. \eqref{eq:repair_limit} limits the total number of restored components in period $k$ to the value of $\boldsymbol{R}_k$, creating a prioritization by enforcing that some components are restored before others. By convention in this work, the value of $\boldsymbol{R}_k$ is set to enforce a consistent number of restorations per time period as follows,
\begin{equation}
\boldsymbol{R}_k =  \left \lceil k \frac{\left| \bcancel{\mathcal{L}} \right|}{\boldsymbol{N}} \right \rfloor \;\; \forall k \in 1,...,\boldsymbol{N}
\end{equation}
Notice that, if the number of damaged lines $\left| \bcancel{\mathcal{L}} \right|$ is equal to the number of restoration periods $N$, then $\boldsymbol{R}_k=k$, and restorations are ordered one at a time. 
If $\left| \bcancel{\mathcal{L}} \right|>N$, the restorations are ordered more coarsely.
We also note that the constraint \eqref{eq:repair_limit} uses $\leq$ instead of strict equality to allow energization to be delayed if required or advantageous.  
Eq. \eqref{eq:repair_active} requires restored components to remain restored for remaining restoration periods.
Finally, Eq. \eqref{eq:repair_all} requires all components to be restored by the final restoration period.

The remaining constraints \eqref{eq:flow_limit_u}-\eqref{eq:power_balance} and \eqref{eq:rip_gen_limits} represent the power flow in the system for each restoration period. 
The power flow $P^L_{ijk}$ and power flow limits on non-damaged lines $\cal L \setminus \bcancel{\cal L}$ are represented 
by \eqref{eq:flow_limit_u} and \eqref{eq:thermal_limit_u}, using a similar formulation as in the RIP problem. 
The line power flow on damaged lines \eqref{eq:flow_limit1_d} and \eqref{eq:flow_limit2_d} use the standard Big-M formulation of line on/off constraints \cite{Burak2016}. A valid Big-M value can found by summing over the maximum angle difference for each branch in the network, then multiplying by the line susceptance, 
\begin{subequations}
\begin{align}
 & \boldsymbol{\theta^{\Delta}} = \sum_{(i,j)\in \mathcal{L}} \overline{\boldsymbol{\theta}_{ij}} \\
& \overline{\boldsymbol{\theta^{\Delta}}_{ij}} = -\underline{\boldsymbol{\theta^{\Delta}}_{ij}} = |\bm b_{ij}|\boldsymbol{\theta^{\Delta}} \;\; \forall (i,j) \in {\cal L}
\end{align}
\end{subequations}

When a line is active, $z^L_{ijk}=1$, these constraints result in the same equality constraint as \eqref{eq:flow_limit_u}. When a line is inactive, $z^L_{ijk}=0$, the line power flow becomes independent of the voltage angles and \eqref{eq:thermal_limit_d} forces the power flow $P^L_{ijk}$ to $0$.
Finally,  \eqref{eq:power_balance} and \eqref{eq:rip_gen_limits} enforce nodal power balance and generation limits, respectively.

Given a solution to the ROP, the associated restoration order $\cal R$ is constructed from the line status variables $z_{ijk}^L$.
The restorations that occur in period $k$, $\mathcal{R}_k$, 
are the lines that transition from inactive to active in period $k$, i.e. if $z_{ij(k)}^L-z_{ij(k-1)}^L=1$ then $\mathcal{R}_k = \langle (i,j)\rangle $.

The scalability of solving the ROP model is limited as the problem size scales with (i) the number of restorations periods $N$, which implies of a larger number of time-coupled problems, and (ii) the network size and number of damaged components considered in each period.
The number of binary decision variables is the number of damaged devices times the number of restoration periods, i.e., to obtain a fully ordered restoration sequence where $N=|\bcancel{\cal L}|$, the number of binary variables scales as $N^2$.
This poor scalability means that even medium sized networks with moderate levels of damage may be impossible to solve with mixed-integer optimization software.  This motivates the need for heuristics that can solve a CRP problem quickly, while obtaining high-quality solutions.

\subsection{Randomized Adaptive Decomposition}

Recognizing the limitations of both UTIL and  ROP approach in solving CRP problems, \cite{coffrin2012last} proposed a meta-heuristic algorithm based on Randomized Adaptive Decomposition (RAD). This algorithm allows for better solution quality than UTIL and better scalability than ROP, and is, to the best of our knowledge, the current state-of-the-art solution algorithm for CRP problems. 
RAD addresses the scalability issue by solving a series of smaller ROP problems on subsets of the restoration sequence.  It requires an initial restoration sequence, which can be obtained for example using UTIL. This sequence is partitioned into small, random subsets with a limited number of restoration periods, and an ROP problem is solved for each subset to create a new restoration ordering. Then a new set of random partitions are created, and the ROP problem is solved again for each subset. The algorithm starts from smaller partitions that are gradually increased in size, and iterates until a time limit is reached.
A more detailed explanation of the algorithm is provided in Appendix \ref{app:RAD}.
Although RAD addresses some of the scalability challenges presented by the ROP, we find that it can take a significant amount of time to converge to high-quality solutions, which limits its practical applicability on large CRPs.

\section{Recursive Restoration Refinement Algorithm} \label{sec:RRR}

The key contribution of this work is the \emph{Recursive Restoration Refinement} algorithm (RRR) for solving CRP problems. RRR is a heuristic method similar to UTIL and RAD that executes a series of two-period ROP problems recursively.
Specifically, the RRR algorithm begins by solving an ROP problem that considers all damaged components in the set $\left| \bcancel{\mathcal{L}} \right|$ and two restoration periods $N=2$, thus coarsely ordering the lines into two sets, where restoration of the first set is prioritized over the second set.  
Each of these sets is then solved by a new two-period ROP problem, increasing the resolution of the restoration ordering to 4 periods. This process continues until the restorations are fully ordered.
In this way, this algorithm builds a tree where the set of restorations are split in half at each stage, until each leaf contains only one restoration action. The final restoration solution is the order of restoration in these leaves.

\begin{figure}[t]
    \begin{center}
    \includegraphics[width=0.48\textwidth]{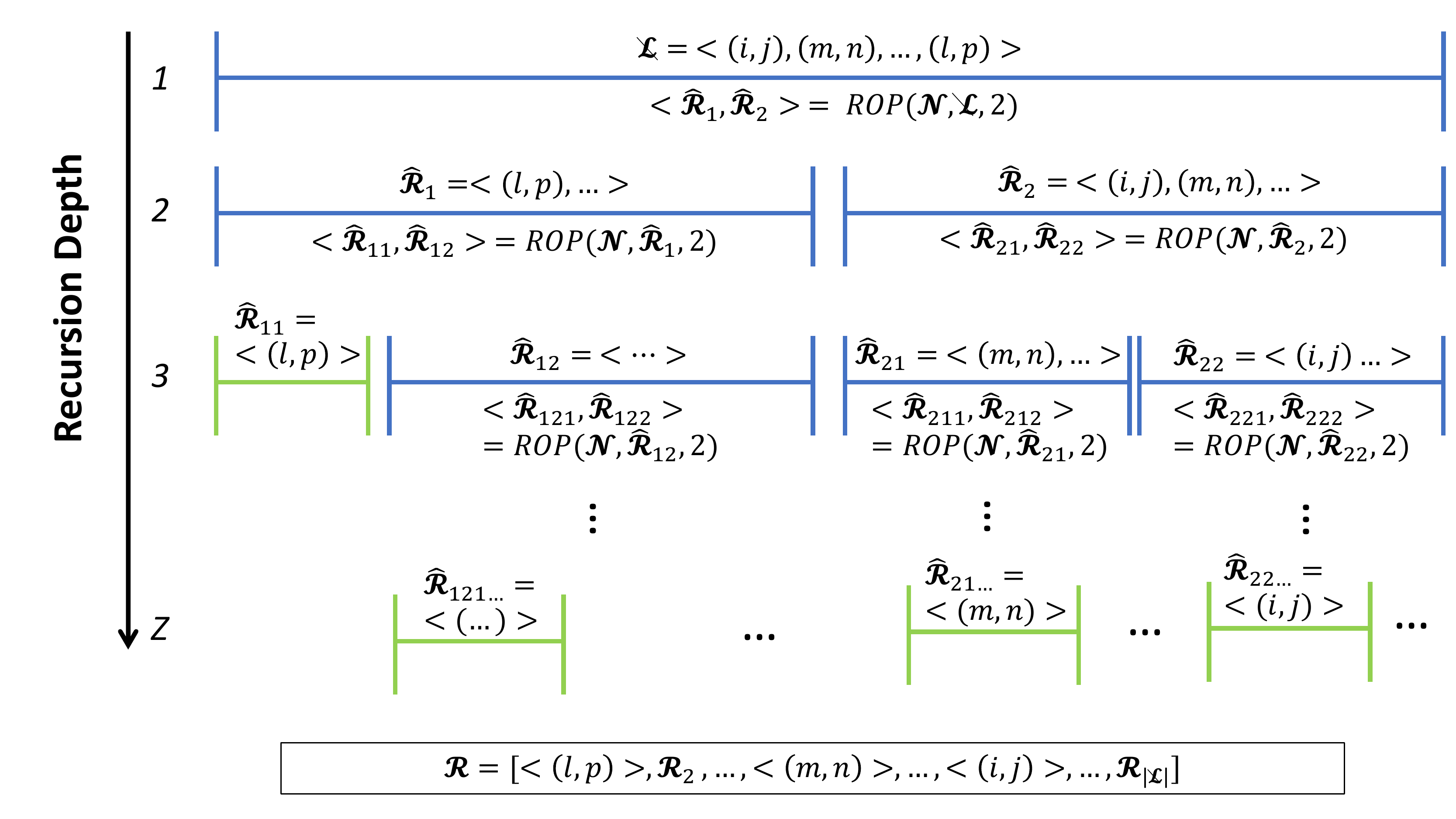}
    \end{center}
    \vspace{-0.3cm}
    \caption{{\small 
    RRR starts with the full set of elements that need restoration (top), and then recursively split the set into two sets, where the first set is prioritized over the other. This process continues until we have a fully ordered restoration sequence with one element restored per period (marked in green). 
    The RRR algorithm thus creates a tree of solutions by solving two-period ROP problems, with the leafs of the tree presenting the final restoration sequence.}}
    \label{fig:RRR_description}
\end{figure}

\begin{algorithm}[t]
    \caption{Recursive Restoration Refinement (RRR)}
    \begin{algorithmic}[1]
        \renewcommand{\algorithmicrequire}{\textbf{Input:}}
        \renewcommand{\algorithmicensure}{\textbf{Output:}}
        \REQUIRE $\mathcal{N}, \bcancel{\mathcal{L}}$ 
        \ENSURE  $\mathcal{R}$\\ 
        \IF {$|\bcancel{\mathcal{L}}| \leq 1$}
            \RETURN [$\bcancel{\mathcal{L}}$] \label{alg:rrr_1}
        \ENDIF
            \STATE $[ \mathcal{\hat R}_1, \mathcal{\hat R}_2 ] \leftarrow ROP(\mathcal{N}, \bcancel{\mathcal{L}}, 2)$ \label{alg:rrr_4}
            \IF {failure($ROP$)}  \label{alg:rrr_5}
                \STATE $\mathcal{R} \leftarrow UTIL(\mathcal{N}, \bcancel{\mathcal{L}})$  \label{alg:rrr_6}
                \STATE $\mathcal{\hat R}_1 \leftarrow \langle \mathcal{R}_k \; \forall k \in 1... \left\lceil  \frac{|\bcancel{\mathcal{L}}|}{2} \right\rceil  \rangle $  \label{alg:rrr_7}
                \STATE $\mathcal{\hat R}_2 \leftarrow \langle\mathcal{R}_k \; \forall k \in \left\lceil \frac{|\bcancel{\mathcal{L}}|}{2} +1\right\rceil ... |\bcancel{\mathcal{L}}| \rangle $  \label{alg:rrr_8}
            \ENDIF
            \IF {$|\mathcal{\hat R}_1| = 0$}  \label{alg:rrr_10}
                \RETURN $UTIL(\mathcal{N}, \bcancel{\mathcal{L}})$  \label{alg:rrr_11}
            \ENDIF
            \STATE $\mathcal{R} \leftarrow [\mathcal{R}_k = \langle \rangle \;\; \forall k \in 1...|\bcancel{\cal L}|]$  \label{alg:rrr_13}
            \STATE $k \leftarrow 1$  \label{alg:rrr_14}
            \FOR {$n \in [1,2]$}
                \STATE $\bcancel{\mathcal{L}}_n \leftarrow \bcancel{\mathcal{L}} \in \mathcal{\hat R}_n$ \label{alg:rrr_16}
                \STATE $ [ \tilde{\mathcal{R}}_1 , \tilde{\mathcal{R}}_2 , \dots , \tilde{\mathcal{R}}_{|\bcancel{\cal L}|}   ]$ $\leftarrow RRR (\mathcal{N}, \bcancel{\mathcal{L}}_n)$ \label{alg:rrr_17}
                \FOR {$l \in [1,2,\dots,{|\bcancel{\cal L}|}]$} \label{alg:rrr_18}
                    \STATE $\mathcal{R}_k \leftarrow \tilde{\mathcal{R}}_l$ \label{alg:rrr_19}
                    \STATE $k \leftarrow k+1$ \label{alg:rrr_20}
                \ENDFOR   \label{alg:rrr_21}
            \ENDFOR
            \RETURN $\mathcal{R}$  \label{alg:rrr_23}
    \end{algorithmic} 
    \label{alg:RRR}
\end{algorithm}

The complete RRR algorithm is shown in Algorithm \ref{alg:RRR}. The inputs are a power grid $\cal N$ and a set of damaged items $\bcancel{\cal L}$. 
The base case is shown in line \ref{alg:rrr_1}. If there is only one damaged component, then no ordering is required and the restoration order $\cal R$ is simply the single damaged component within $\bcancel{\cal L}$. If there is more than 1 damaged component a 2 period ROP problem is solved on line \ref{alg:rrr_4}. This returns two restoration periods $\mathcal{\hat R}_1$ and $\mathcal{\hat R}_2$, with components in $\mathcal{\hat R}_1$ being prioritized over components in $\mathcal{\hat R}_2$.  
There are two special cases that must be considered before continuing with the next recursion step.  

The first case is when the two-period ROP problem fails to find a solution, for example, as a result of a numerical error or a time limit (line \ref{alg:rrr_4}).  If this situation occurs, the UTIL heuristic is used to split the repairs into two periods (lines \ref{alg:rrr_5}-\ref{alg:rrr_8}).

The second case is when no restorations occur in the first period (line \ref{alg:rrr_10}). This suggests there are no components that are more urgent than others to restore, which occurs, e.g., when power is fully restored. In this case, any ordering works equally well and for convenience, UTIL is again used to order the repairs (line \ref{alg:rrr_11}).
When the UTIL algorithm is used, RIP is also solved to calculate power flow for these restoration periods.

The primary recursive procedure occurs in lines \ref{alg:rrr_13}-\ref{alg:rrr_23}, where the RRR problem is solved using the same grid $\cal N$ on the subset of components occurring in the two restoration periods $\mathcal{\hat R}_1$ and $\mathcal{\hat R}_2$ (line \ref{alg:rrr_17}).
The final restoration order $\cal R$ is constructed by merging the outputs of these recursive RRR calls in lines \ref{alg:rrr_18}-\ref{alg:rrr_21}.

The key value of the RRR algorithm is its ability to scale to much larger problems than traditional ROP.
This is possible because the ROP sub-problems posed by RRR always have exactly $2$ periods (i.e., $N=2$). Therefore, the largest ROP sub-problem that is solved by RRR has $2|\bcancel{\cal L}|$ binary variables, in contrast to the $|\bcancel{\cal L}|^2$ variables required by the traditional ROP formulation.
However, RRR is only a heuristic solution and does not guarantee optimality. Further, solving the two-period ROP may still be time consuming. 
However the experimental results will show that in practice the RRR algorithm provides an ideal tradeoff of runtime requirements and solution quality.

\section{Post-Processing of Solutions}
\label{sec:processing}
All of the algorithms presented above provide a restoration order $\mathcal{R}$. The ROP, RAD and RRR problem also provides information about the total power served in each time step, and similar information can be obtained for the UTIL restoration order by solving the RIP problem. 
However, in our simulations, we observe that the total amount of power served in each time step can sometimes drop at intermediate periods during the restoration process.
This counter intuitive effect is known as {\em Brace's Paradox} \cite{blumsack2006braess,9127501}, where increasing the total line capacity in the networks decreases the effective capacity. 
This effect can occur with the UTIL and RRR algorithms as their restoration orders $\mathcal{R}$ always have one component being restored in each time step.  It can also occur in ROP and RAD if the ROP problem is not fully solved to optimality. 

This reduction in power delivery is an artifact of the mathematical modeling of this work, as any practical CRP solution would delay re-energization of additional components until it is possible to avoid a drop in the served load.
Hence, we remove the effects of Brace's Paradox by post processing the amount of power served in each restoration period $k$ to be the maximum power served in any previous period, i.e.,
\begin{equation}
    \textstyle{P^{\text{Total}}_k=\max_{n=1,\dots,k}\left\{\sum_{\demandid \in {\demand}} x^D_{\demandid n}   \boldsymbol{P}^D_{\demandid} \boldsymbol{\Delta}_n\right\}} 
    \nonumber
\end{equation}
This post-processing achieves the effect of delaying the activation of components in a period $\mathcal{R}_k$ until they increase power delivery.  Formally, if the power delivered is reduced in the periods $k$ to $k+m$, then the devices restored in those periods are instead restored in the period $k+m+1$, and the restoration set for periods in $k$ to $k+m$ are empty,
$$\mathcal{R}_{k+m+1} = Union(\mathcal{R}_{k},...,\mathcal{R}_{k+m+1})$$
$$\mathcal{R}_n = \langle \rangle \quad \forall n \in k,...,k+m$$
This post processing is performed on all solutions. 

In addition to the post-processing discussed above, it would be possible to perform a post-processing of the restoration ordering solutions to assess AC power flow feasibility. Specifically, we could take the restoration ordering obtained from our algorithms and solve an AC optimal power flow at each time step to assess whether the proposed restoration sequence is AC feasible. Based on \cite{rhodes2021powermodelsrestoration}, solving an AC optimal power flow (which allows for redispatch of generation, even if the restoration sequence is fixed) is likely to produce feasible results on small networks, while simply running an AC power flow simulation often leads to infeasibility and non-convergence \cite{6345338}. However, in this paper, we omit an analysis of AC power flow feasibility because of space limitations. 

\section{Results}\label{sec:results}

To study the efficacy of the different restoration algorithms, we select six power systems from the IEEE Power Grid OPF Library v21.07 \cite{babaeinejadsarookolaee2019power} that range from 24 to 500 buses, which are listed in Table \ref{tab:case_study}.  We use the API grid models, which have increased active power injections, to ensure the loading of the grids is high and the optimization problems are challenging cases for restoration ordering. 

Damage to the systems was applied to randomly selected transmission lines, with the total damage count ranging from 10\% to 100\% percent of lines. 
Table \ref{tab:case_study} shows the number of damaged lines for each test case and damage percentage.  
Overall, each of the restoration algorithms are evaluated on 60 distinct damage scenarios. 

\begin{table}
    \centering
    \caption{\small Selected power system API test cases from PGLib. Highlighted scenarios are solved within an optimality gap of 1\% by ROP.}
\resizebox{\columnwidth}{!}{%
\begin{tabular}{r|r|r|r|r|r|r|r|r|r|r|}
\cline{2-11}
\multicolumn{1}{c|}{} & \multicolumn{10}{c|}{\textbf{Damaged Line Count}} \\ \hline 
\multicolumn{1}{|c|}{\textbf{Case}} &\textit{10\%}&\textit{20\%}&\textit{30\%}&\textit{40\%}&\textit{50\%}&\textit{60\%}&\textit{70\%}&\textit{80\%}&\textit{90\%}&\textit{100\%}\\ \hline \hline
\multicolumn{1}{|r|}{\textit{case24}}  & \cellcolor[HTML]{C0C0C0} \textbf{4} & \cellcolor[HTML]{C0C0C0} \textbf{8} & \cellcolor[HTML]{C0C0C0} \textbf{11} & \cellcolor[HTML]{C0C0C0} \textbf{15} & \cellcolor[HTML]{C0C0C0} \textbf{19} & \cellcolor[HTML]{C0C0C0} \textbf{23} & \cellcolor[HTML]{C0C0C0} \textbf{27} & \cellcolor[HTML]{C0C0C0} \textbf{30} & \cellcolor[HTML]{C0C0C0} \textbf{34} & \cellcolor[HTML]{C0C0C0} \textbf{38}  \\ \hline
\multicolumn{1}{|r|}{\textit{case39}}  & \cellcolor[HTML]{C0C0C0} \textbf{5} & \cellcolor[HTML]{C0C0C0} \textbf{9} & \cellcolor[HTML]{C0C0C0} \textbf{14} & \cellcolor[HTML]{C0C0C0} \textbf{18} & \cellcolor[HTML]{C0C0C0} \textbf{23} & \cellcolor[HTML]{C0C0C0} \textbf{28} & \cellcolor[HTML]{C0C0C0} \textbf{32} & \cellcolor[HTML]{C0C0C0} \textbf{37} & \cellcolor[HTML]{C0C0C0} \textbf{41} & \cellcolor[HTML]{C0C0C0} \textbf{46}  \\ \hline
\multicolumn{1}{|r|}{\textit{case60}}  & \cellcolor[HTML]{C0C0C0} \textbf{9} & \cellcolor[HTML]{C0C0C0} \textbf{18} & \cellcolor[HTML]{C0C0C0} \textbf{26} & \cellcolor[HTML]{C0C0C0} \textbf{35} & \cellcolor[HTML]{C0C0C0} \textbf{44} & \cellcolor[HTML]{C0C0C0} \textbf{53} & \cellcolor[HTML]{C0C0C0} \textbf{62} & \cellcolor[HTML]{C0C0C0} \textbf{70} & \cellcolor[HTML]{C0C0C0} \textbf{79} & \cellcolor[HTML]{C0C0C0} \textbf{88}  \\ \hline
\multicolumn{1}{|r|}{\textit{case118}}  & \cellcolor[HTML]{C0C0C0} \textbf{19} & \cellcolor[HTML]{C0C0C0} \textbf{37} & \cellcolor[HTML]{C0C0C0} \textbf{56} & \cellcolor[HTML]{C0C0C0} \textbf{74} & \cellcolor[HTML]{C0C0C0} \textbf{93} & 112 & 130 & 149 & 167 & 186  \\ \hline
\multicolumn{1}{|r|}{\textit{case240}}  & \cellcolor[HTML]{C0C0C0} \textbf{45} & \cellcolor[HTML]{C0C0C0} \textbf{90} & \cellcolor[HTML]{C0C0C0} \textbf{134} & \cellcolor[HTML]{C0C0C0} \textbf{179} & \cellcolor[HTML]{C0C0C0} \textbf{224} & 269 & 314 & 358 & 403 & 448  \\ \hline
\multicolumn{1}{|r|}{\textit{case500}}  & \cellcolor[HTML]{C0C0C0} \textbf{73} & \cellcolor[HTML]{C0C0C0} \textbf{147} & 220 & 293 & 366 & 440 & 513 & 586 & 660 & 733  \\ \hline
\end{tabular}
}
    \label{tab:case_study}
\end{table}

\subsection{Implementation}
The optimization problems and algorithms are implemented in \emph{PowerModelsRestoration} \cite{rhodes2021powermodelsrestoration}, an open source package in the Julia programming language \cite{julia}.  All optimization problems are solved using the Gurobi v9.1 solver \cite{gurobi}. All problems are solved on a  2 cpu  server with 36 cores running at 2.1 GHz, and 128 GB of memory. 

We next discuss the parameters used for implementation of the mixed-integer problems involved in the ROP, RAD and RRR. (UTIL is not discussed, as it does not require the definition of any particular parameters.)

\begin{itemize}
    \item \emph{Optimality gap:} We solve the mixed-integer problems involved in ROP, RAD, and RRR with a 1\% optimality gap, which significantly reduced the solution times relative to the default .01\% gap.  
We also tried a 5\% and 10\% optimality gap, but these higher values resulted in a severe reduction in solution quality.

\item \emph{Time limit:} 
We applied different time limits for the methods, ranging from 10 hours to 5 minutes. 

\item \emph{Time step duration:} The duration of $\mathbf{\Delta}_k$ is 1 hour for simplicity.
\end{itemize}

In addition to these shared parameters, we also include some specific parameters for each algorithm:
\begin{itemize}
    \item \emph{Restoration Ordering Problem:} 
    We warm-start the ROP with the restoration ordering from UTIL to help the solver find an initial feasible solution. 
    (Note that warmstarting is not required for RAD or RRR as they are able to find feasible solutions without a warm start.)

\item\emph{Randomized Adaptive Decomposition Algorithm:}
Following the design proposed in \cite{coffrin2012last}, this approach begins with small partitions of the complete restoration order (i.e., 2-5 components each) with tight time limits (1\% of the algorithm's full time limit).  As the algorithm progresses the partition sizes and sub-problem time limits are increased to find higher quality solutions as explained in the appendix.
The algorithm terminates when it reaches the algorithm time limit, or after 100 iterations within which the solution did not improve. 

\item \emph{Recursive Restoration Refinement Algorithm:}
The sub-ROP problem is assigned a time limit equal to half the remaining algorithm time $t_{ROP} = t_{RRR}/2 $. This time constraint provides the most time for the first problem that splits all of the restorations, but leaves sufficient time for remaining problems in the recursive call to calculate a solution.
\end{itemize}

In the rest of this section,
we examine the solution quality and solution time of the restoration algorithms.  

\subsection{Benchmarking against ROP Optimal Solutions}
We first consider solutions to instances where the ROP obtains a solution with $<1$\% optimality gap within 10 hours. In this case, the ROP solution is provably optimal (up to the tolerance), and we use this solution to benchmark the solution quality of the heuristic algorithms.

\subsubsection{Optimal Solution for a Single Case}
\begin{figure}[t]
    \centering
    \begin{subfigure}[t]{0.24\textwidth}
        \centering
        \includegraphics[width=1.0\textwidth]{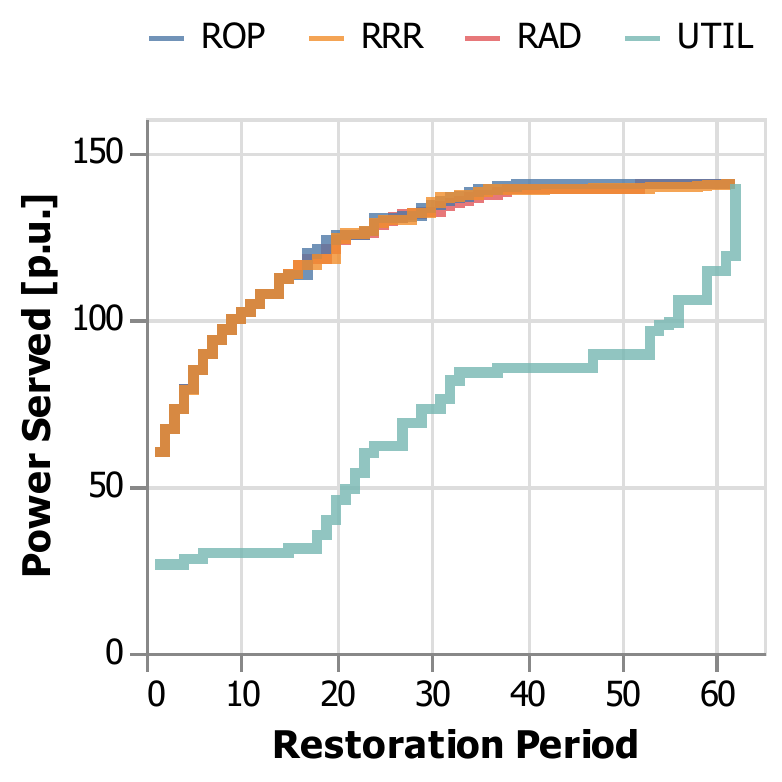}
        \caption{\footnotesize \textbf{Case60 with 70\% Line Damage} 
        } \label{fig:duration_plot_optimal}
    \end{subfigure}    
    \begin{subfigure}[t]{0.24\textwidth}
        \centering
        \includegraphics[width=1.0\textwidth]{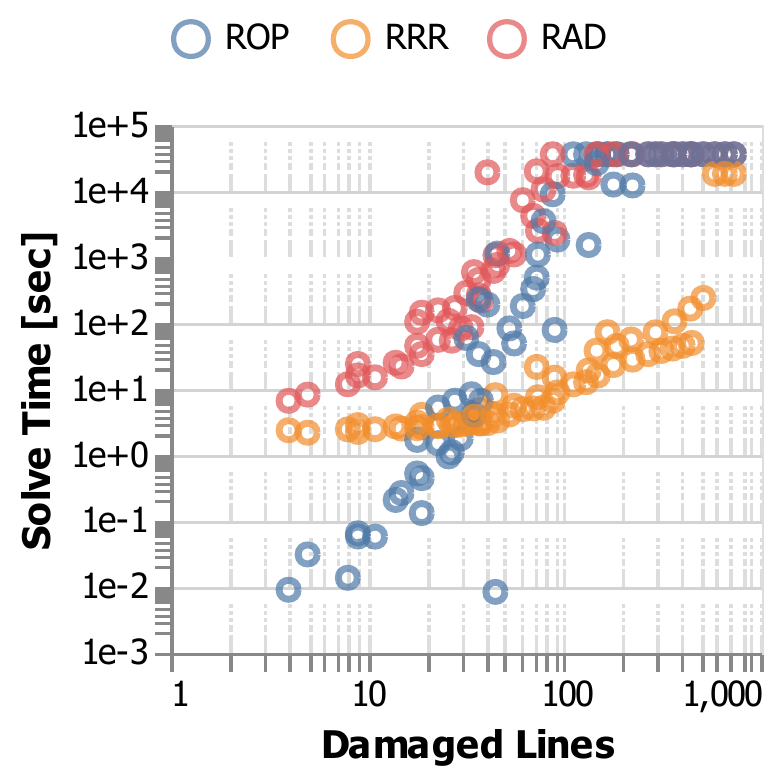} 
        \caption{\footnotesize \textbf{Algorithm Scaling}
        } \label{fig:long_scatter}
    \end{subfigure}
    \caption{\small 
    \textbf{(a)} Power served for each restoration period. ROP, RAD, and RRR at first restores load quickly, then slower. UTIL does not show a similar pattern. 
    \textbf{(c)} The algorithm solve time increases with a larger number of damaged lines.  RRR scales notably better than ROP or RAD, finding solutions up 1,000 times faster. 
    } \label{fig:long}
    \vspace{-10pt}
\end{figure}

We first look at the load served over time for case60 with 70\% damage to lines, which ROP solves to optimality. Figure \ref{fig:duration_plot_optimal} shows the load served over time by the different algorithms.  The total load served, which can be interpreted as the integral under the curves, is 7753.1 MW for ROP, 7713.4 MW for RRR, 7712.6 MW for RAD and 4227.1 MW for UTIL. 

This solution is typical for cases when ROP is able to solve to optimality. The total load served by ROP, RAD, and RRR is within the 1\% optimality gap and the load curves for the three algorithms are very similar, though not exactly the same. 
In contrast, the UTIL solution serves much less load  and often fails to improve the load served for several restoration periods. 

\subsubsection{Quality of solution across multiple cases}
Total power served for other scenarios solved with a 10-hour time limit are shown in Table \ref{tab:ENS_10h_data}, where the best solution for each grid scenario and solutions within 1\% of the best solutions are highlighted and marked in bold font and a darker shade. 
We consider the subset of 42 cases that ROP solves to optimality, which are also indicated by a darker shade and bold font in Table \ref{tab:case_study}, and calculate the mean and standard deviation of percentage of power served across those 42 cases.
We observe that the ROP solutions serve on average 93.6\% of power demand, while RRR serves 93.4\%  and RAD serves 92.1\%.  In contrast, UTIL is only able to serve 77.2\% of power demand on average. 
Further, the standard deviation of percent power served for ROP, RRR, and RAD are around 0.06\%, meaning that most solutions are very close to the (high) average values. 
In comparison, the standard deviation for UTIL is 0.195, which indicates a much higher variability in solution quality.

We conclude from these results that RRR generally obtains solutions that are very close to the optimal ROP solutions, typically slightly better than the RAD solutions and much better than the UTIL solutions.

\subsection{Algorithm Run Time Requirements}
To analyze the scalability of the algorithm, we plot the number of damaged components and solution time for each of our cases, for each level of damage. We again use results obtained with an upper time limit of 10 hours, and note that not all solutions points are optimal.
Fig. \ref{fig:long_scatter} shows the number of damaged lines in each scenario (horizontal axis) vs the time needed to solve each case (vertical axis) for ROP, RRR, and RAD.   
For scenarios with more than 100 damaged lines, ROP and RAD both terminate at the 10h time limit, while RRR typically terminates within 10-100s. This means that RRR frequently is 1,000 times faster than ROP and RAD. Only the three largest cases take a long time to solve RRR, because the problem at recursion depth 1 is not solved to optimality within the allocated time of 5 hours.

\subsection{Benchmarking on Time-Limited Solutions}
\begin{figure}
\centering
    \begin{subfigure}{0.24\textwidth}
        \centering
        \includegraphics[width=1.0\textwidth]{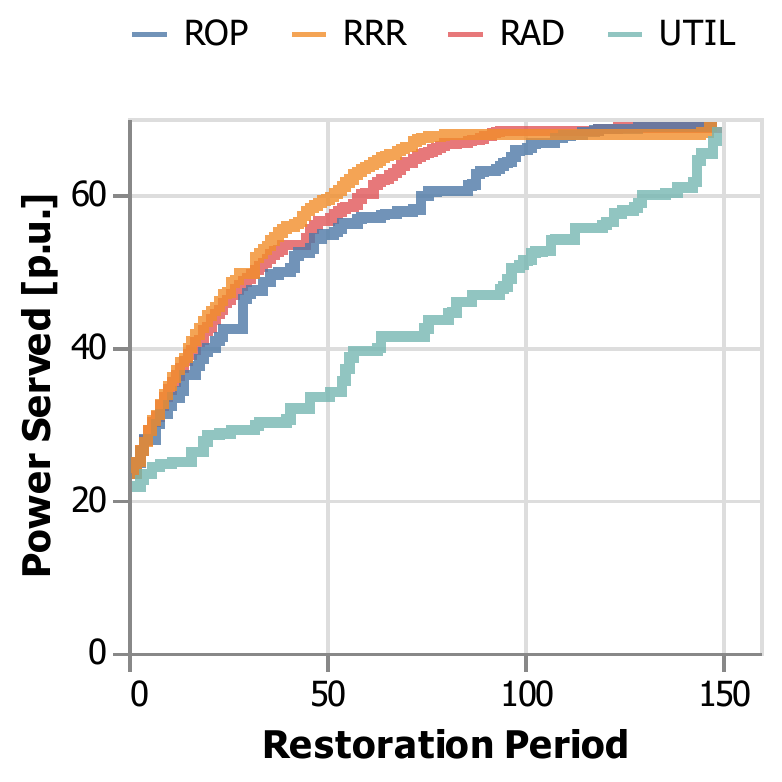}
        \caption{\footnotesize \textbf{Power Served Over Time}
        } \label{fig:speed_duration}
    \end{subfigure}
    \begin{subfigure}{0.24\textwidth}
        \centering
        \includegraphics[width=1.0\textwidth]{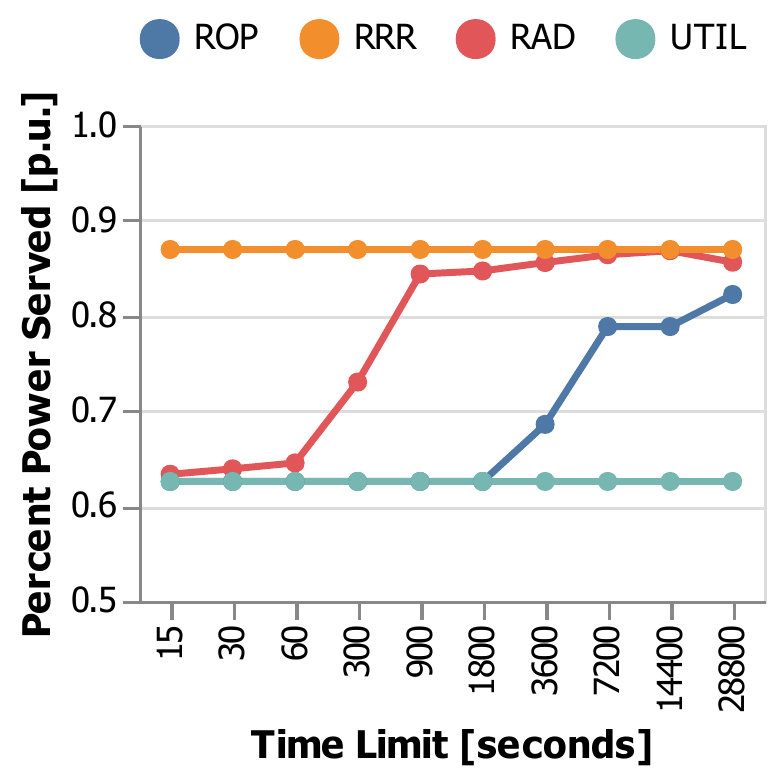} 
        \caption{\footnotesize \textbf{Performance vs Time Limit} 
        } \label{fig:performance_profile}
     \end{subfigure}
    \caption{\small Results for case118 with 80\% line damage.
    \textbf{(a)} Power served for each restoration periods. With 10 hours ROP cannot find an optimal solution, and both RRR and RAD serve more power than ROP.
    \textbf{(b)} Total power served versus the algorithm time limit.
    }
     \label{fig:short}
\end{figure}

The case118 with 80\% damage (corresponding to 149 damaged lines) is one of the cases where ROP was unable to find an optimal solution even in 10 hours. The power served over time for this scenario is shown in Fig. \ref{fig:speed_duration}. We observe that the RRR is the best solution in this case, with a total power served of 8911.1 MW. RAD is almost as good with 8792.9 MW served, while ROP only serve 8426.1 MW of power.  UTIL is the worst solution with 6411.3 MW power served. 
Based on those results and the plots, we conclude that both RAD and RRR outperform UTIL and ROP in this case.

\subsubsection{Performance Profile: Case 118 80\%}
To understand how solution quality changes as more time is made available to the algorithm,
we solve case118 with 80\% line damage several times while varying the time limit between 15 seconds to 8 hours. Fig. \ref{fig:performance_profile} plots the percent of total demand served versus the algorithm time limit. The UTIL and RRR algorithms both solve the problem within 15 seconds, and thus do not improve with more time.  ROP and RAD are both initialized with the Utilization solution.  RAD begins improving on that solution within 15 seconds, but requires 15 minutes to approach the same solution quality as RRR and 2 hours to come close to it.  ROP begins to improve the solution after 1 hour and continues to improve until the time limit is reached, but is still not able to match the RRR or RAD solutions. 

\subsubsection{Solution Quality with 5 Minute Time Limit}
We next solve each scenario with a five minute algorithm time limit.  
We choose this time limit because we believe that it represents a time frame that is short enough for the system operator to solve (and resolve) the restoration ordering problem as the restoration progresses. Particularly in situations where the overall state of the system may initially be unknown, it is important to be able to resolve the problem quickly as new information arrives.

The full results are shown in Table \ref{tab:ENS_5m_data}, where the best solution for each grid scenario and solutions within 1\% of the best solutions are highlighted and marked in bold font.
RRR is consistently either the best solution or within 1\% of the best solution for all cases, even for small cases with low number of damaged items where ROP solves to optimality. RAD outperforms ROP on large systems, but it rarely matches the solution quality of RRR.  

\begin{figure*}[t]
    \centering
    \begin{subfigure}[t]{0.32\textwidth}
        \centering
        \includegraphics[width=1.0\textwidth]{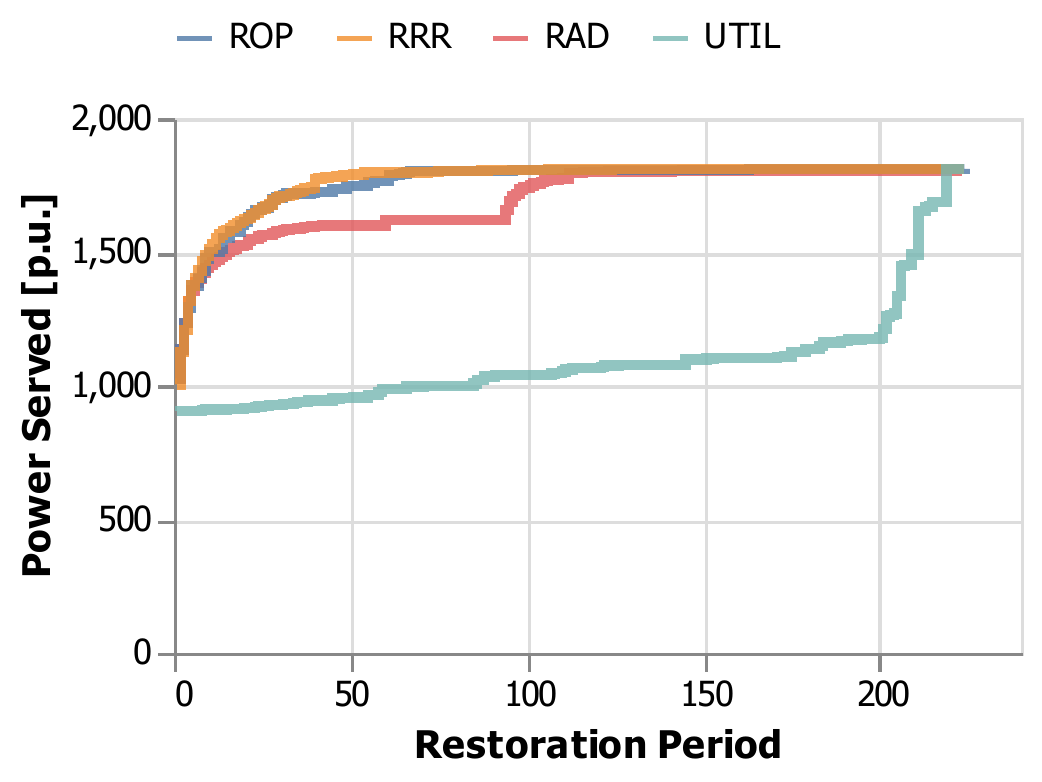}
        \caption{\footnotesize \textbf{Power Served}} \label{fig:analysis_duration}
    \end{subfigure}
    \begin{subfigure}[t]{0.32\textwidth}
        \centering
        \includegraphics[width=1.0\textwidth]{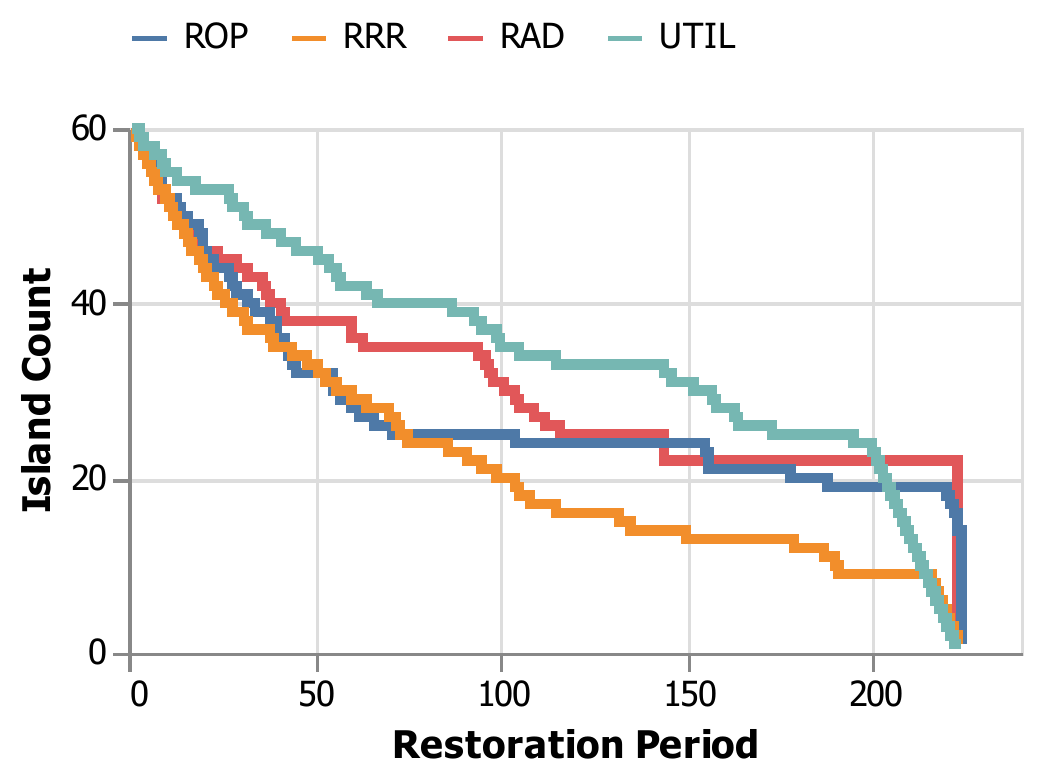} 
        \caption{\footnotesize \textbf{Island Count}} \label{fig:analysis_island_count}
     \end{subfigure}
     \begin{subfigure}[t]{0.32\textwidth}
        \centering
        \includegraphics[width=1.0\textwidth]{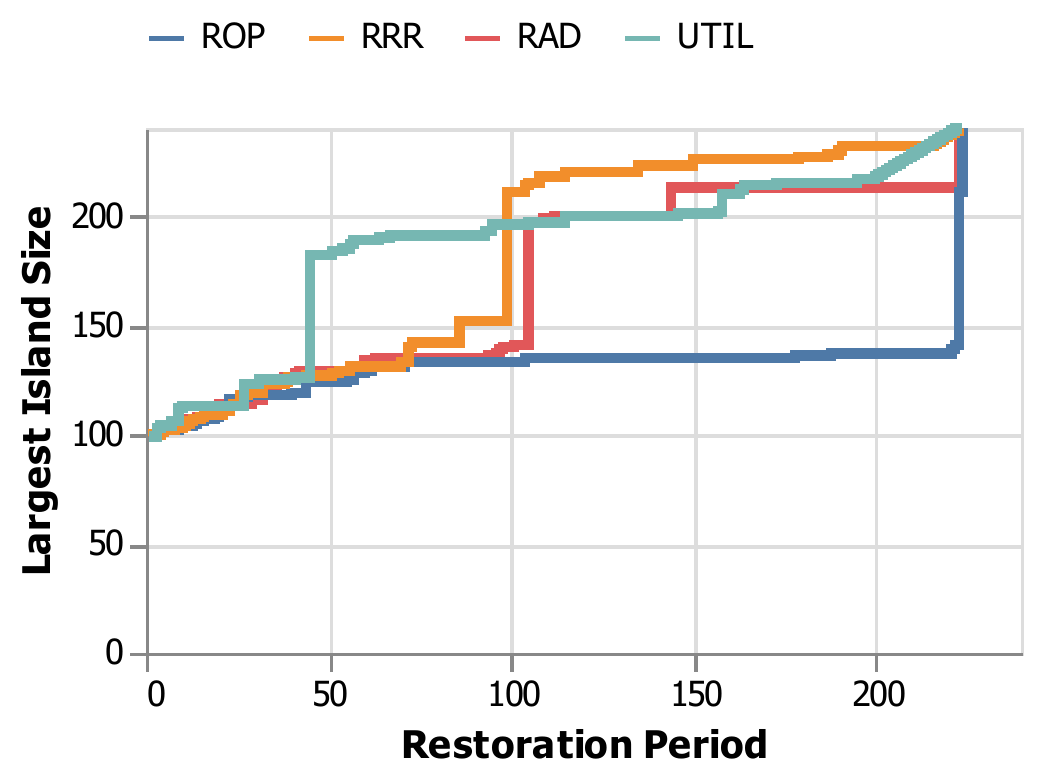} 
        \caption{\footnotesize \textbf{Size of Largest Island}} \label{fig:analysis_largest_island}
    \end{subfigure}
    \caption{\small Restoration solutions for case240 with 50\% line damage.
    \textbf{(a)} Power served vs restoration period. The RRR and RAD algorithms intially quickly increases the load, and then slow down. The RAD and UTIL algorithms follow a less clear pattern.
    \textbf{(b)} Island count vs restoration period. The RRR and ROP algorithms quickly reduce the number of islands in the network, while RAD and UTIL maintain more islands.
    \textbf{(c)} Size of largest island. Quickly creating a large island is not essential for good solutions.
    }
     \label{fig:analysis}
\end{figure*} 

Comparing the statistics of the 5-min. solutions across all cases, we observe that the RRR solutions serve on average 91.9\% of power demand with a low standard deviation of 0.06\%, which is comparable to the results obtained with a 10-hour limit. ROP and RAD both serve an average of 78.0\% of the power demand, which is much worse than for solutions obtained with a 10-hour time limit, while UTIL still serves {68.3\%} of power demand on average. The standard deviation of ROP, RAD, and UTIL solutions are 0.271, 0.230, and {0.242}, respectively, indicating that the solution quality varies widely but is generally low.

\subsection{Analyzing Restoration Results}
Many factors affect the optimal CRP solution including the number of damaged items, connectivity between loads and generators, and network characteristics like line capacity and redundancy. Although these factors are complex, we attempt to draw some general conclusions about the characteristics of high-quality solutions to the CRP problem. By investigating the whole range of restoration solutions, three trends stand out.
\begin{enumerate}
    \item Good solutions initially restore load quickly before gradually slowing down. 
    \item It is important to quickly reduce the number of islands to connect load and generation resources.
    \item Increasing the size of the largest island is not essential.
\end{enumerate}

To illustrate these effects on a particular example, 
Fig. \ref{fig:analysis_duration} shows the CRP solutions for each algorithm on the Case 240 system with 50\% damage to lines, given a 10h time limit.
We show the total amount of power served increases with the restoration period in Fig. \ref{fig:analysis_duration}, the number of islands in the system in Fig. \ref{fig:analysis_island_count} and the size of the largest island in Fig. \ref{fig:analysis_largest_island}.

First, we observe from Fig. \ref{fig:analysis_duration} that the RRR algorithm finds the best solution (i.e., the solution with the highest power served), closely followed by ROP.  RAD has a slightly worse solution, and UTIL is poor in comparison. 
Fig. \ref{fig:analysis_island_count} shows that the RRR and ROP solutions rapidly decrease the number of islands, until almost all power demand is served in restoration period 75.  The RAD solution follows a similar trajectory until restoration period 25 (where it stops performing well), while the UTIL algorithm does not reduce the number of islands nearly as quickly as the other algorithms.
When considering the largest island plotted in Fig. \ref{fig:analysis_largest_island},
we observe that the UTIL algorithm creates a large grid quickly as it connects the largest capacity lines first.  In comparison, RRR, ROP, and RAD delay forming a very large grid until after they fully serve load. Instead, these solutions first focus on creating smaller networks to power all loads, before interconnecting.  
This result suggests that future work may need to consider additional details regarding the process of reconnecting multiple islands during restoration, such as dynamic stability and need for resynchronization.

\section{Conclusion}\label{sec:conclusion}
In this paper, we propose the Recursive Restoration Refinement algorithm to identify effective restoration sequences that maximize the delivered power in power systems with a large number of damaged lines. The algorithm recursively orders the repairs into two halves, where the first half is prioritized over the second, until the repairs are organized into a restoration sequence with one repair per period. 
This algorithm finds near-optimal solutions 1000 times faster than existing algorithms. The solutions generally fall within 1\% of the provably optimal solution obtained by solving the Restoration Ordering Problem. On larger test cases, we are able to identify good restoration sequences for large systems (tested up to 500 nodes) and a large number of damaged components (tested for up to 700 damaged lines) in less than five minutes. In comparison, the Restoration Ordering Problem terminates with a very sub-optimal solution after 10 hours. 

Future work will expand the model to consider damage to more components, such as buses and generators, and consider
non-linear power flow formulations to be used, such as AC power flow or convex relaxations.
Further, the RRR algorithm may replace UTIL as a warm-start solution for ROP or RAD, and could be combined with other pre-processing tools to further improve performance.
Overall, our goal is to extend this work and successfully order 1,000 repairs on a 5,000 bus network, with an AC-feasible power flow solution. 

\bibliographystyle{IEEEtran}
\bibliography{IEEEabrv,references.bib}

\appendix

\begin{table*}[t]
\centering
\caption{ Total Power Served [p.u.] with 10 hour algorithm time limit.}
\begin{tabular}{|l||r|r|r|r|r|r|r|r|r|r|}
\hline 
\textbf{Case} & \multicolumn{10}{c|}{\textbf{case24api}}\\ \hline 
\textbf{Damage} & \multicolumn{1}{c|}{\textbf{10}}& \multicolumn{1}{c|}{\textbf{20}}& \multicolumn{1}{c|}{\textbf{30}}& \multicolumn{1}{c|}{\textbf{40}}& \multicolumn{1}{c|}{\textbf{50}}& \multicolumn{1}{c|}{\textbf{60}}& \multicolumn{1}{c|}{\textbf{70}}& \multicolumn{1}{c|}{\textbf{80}}& \multicolumn{1}{c|}{\textbf{90}}& \multicolumn{1}{c|}{\textbf{100}}\\ \hline 
\textbf{Power Demand} & \multicolumn{1}{c|}{\textbf{218.8}}& \multicolumn{1}{c|}{\textbf{437.6}}& \multicolumn{1}{c|}{\textbf{601.7}}& \multicolumn{1}{c|}{\textbf{820.6}}& \multicolumn{1}{c|}{\textbf{1039.4}}& \multicolumn{1}{c|}{\textbf{1258.2}}& \multicolumn{1}{c|}{\textbf{1477.0}}& \multicolumn{1}{c|}{\textbf{1641.1}}& \multicolumn{1}{c|}{\textbf{1859.9}}& \multicolumn{1}{c|}{\textbf{2078.8}}\\ \hline 
\emph{Util}& 216.5 & \cellcolor[HTML]{C0C0C0} \textbf{434.5} & 590.7 & 771.0 & 918.0 & 1095.0 & 1319.1 & 1353.6 & 1410.9 & 1521.8 \\ \hline 
\emph{RAD}& \cellcolor[HTML]{C0C0C0} \textbf{218.8} & \cellcolor[HTML]{C0C0C0} \textbf{435.8} & \cellcolor[HTML]{C0C0C0} \textbf{597.2} & \cellcolor[HTML]{C0C0C0} \textbf{801.1} & \cellcolor[HTML]{C0C0C0} \textbf{998.0} & \cellcolor[HTML]{C0C0C0} \textbf{1192.8} & \cellcolor[HTML]{C0C0C0} \textbf{1428.4} & 1487.4 & 1579.5 & \cellcolor[HTML]{C0C0C0} \textbf{1845.8} \\ \hline 
\emph{RRR}& \cellcolor[HTML]{C0C0C0} \textbf{218.8} & \cellcolor[HTML]{C0C0C0} \textbf{436.0} & \cellcolor[HTML]{C0C0C0} \textbf{596.8} & \cellcolor[HTML]{C0C0C0} \textbf{801.1} & \cellcolor[HTML]{C0C0C0} \textbf{1003.0} & \cellcolor[HTML]{C0C0C0} \textbf{1194.6} & \cellcolor[HTML]{C0C0C0} \textbf{1426.1} & \cellcolor[HTML]{C0C0C0} \textbf{1524.8} & \cellcolor[HTML]{C0C0C0} \textbf{1668.6} & \cellcolor[HTML]{C0C0C0} \textbf{1860.2} \\ \hline 
\emph{ROP}& 216.5 & \cellcolor[HTML]{C0C0C0} \textbf{434.5} & \cellcolor[HTML]{C0C0C0} \textbf{596.0} & \cellcolor[HTML]{C0C0C0} \textbf{800.4} & \cellcolor[HTML]{C0C0C0} \textbf{1000.7} & \cellcolor[HTML]{C0C0C0} \textbf{1200.4} & \cellcolor[HTML]{C0C0C0} \textbf{1424.8} & \cellcolor[HTML]{C0C0C0} \textbf{1530.5} & \cellcolor[HTML]{C0C0C0} \textbf{1679.4} & \cellcolor[HTML]{C0C0C0} \textbf{1864.4} \\ \hline 
\hline 
\textbf{Case} & \multicolumn{10}{c|}{\textbf{case39api}}\\ \hline 
\textbf{Damage} & \multicolumn{1}{c|}{\textbf{10}}& \multicolumn{1}{c|}{\textbf{20}}& \multicolumn{1}{c|}{\textbf{30}}& \multicolumn{1}{c|}{\textbf{40}}& \multicolumn{1}{c|}{\textbf{50}}& \multicolumn{1}{c|}{\textbf{60}}& \multicolumn{1}{c|}{\textbf{70}}& \multicolumn{1}{c|}{\textbf{80}}& \multicolumn{1}{c|}{\textbf{90}}& \multicolumn{1}{c|}{\textbf{100}}\\ \hline 
\textbf{Power Demand} & \multicolumn{1}{c|}{\textbf{505.3}}& \multicolumn{1}{c|}{\textbf{909.6}}& \multicolumn{1}{c|}{\textbf{1415.0}}& \multicolumn{1}{c|}{\textbf{1819.2}}& \multicolumn{1}{c|}{\textbf{2324.6}}& \multicolumn{1}{c|}{\textbf{2829.9}}& \multicolumn{1}{c|}{\textbf{3234.2}}& \multicolumn{1}{c|}{\textbf{3739.6}}& \multicolumn{1}{c|}{\textbf{4143.8}}& \multicolumn{1}{c|}{\textbf{4649.2}}\\ \hline 
\emph{Util}& 451.5 & 843.3 & 1193.4 & 1254.6 & 1463.5 & 1612.4 & 1731.6 & 1644.9 & 1810.7 & 1633.1 \\ \hline 
\emph{RAD}& 484.0 & \cellcolor[HTML]{C0C0C0} \textbf{877.3} & \cellcolor[HTML]{C0C0C0} \textbf{1332.6} & 1608.0 & 1970.3 & \cellcolor[HTML]{C0C0C0} \textbf{2490.1} & \cellcolor[HTML]{C0C0C0} \textbf{2721.6} & 3129.4 & \cellcolor[HTML]{C0C0C0} \textbf{3439.2} & 3661.4 \\ \hline 
\emph{RRR}& \cellcolor[HTML]{C0C0C0} \textbf{492.5} & \cellcolor[HTML]{C0C0C0} \textbf{876.3} & \cellcolor[HTML]{C0C0C0} \textbf{1333.5} & \cellcolor[HTML]{C0C0C0} \textbf{1669.4} & \cellcolor[HTML]{C0C0C0} \textbf{1994.5} & \cellcolor[HTML]{C0C0C0} \textbf{2490.7} & \cellcolor[HTML]{C0C0C0} \textbf{2694.5} & \cellcolor[HTML]{C0C0C0} \textbf{3133.1} & \cellcolor[HTML]{C0C0C0} \textbf{3414.4} & \cellcolor[HTML]{C0C0C0} \textbf{3667.7} \\ \hline 
\emph{ROP}& \cellcolor[HTML]{C0C0C0} \textbf{488.4} & \cellcolor[HTML]{C0C0C0} \textbf{876.7} & \cellcolor[HTML]{C0C0C0} \textbf{1339.1} & \cellcolor[HTML]{C0C0C0} \textbf{1668.6} & \cellcolor[HTML]{C0C0C0} \textbf{2011.2} & \cellcolor[HTML]{C0C0C0} \textbf{2511.3} & \cellcolor[HTML]{C0C0C0} \textbf{2721.2} & \cellcolor[HTML]{C0C0C0} \textbf{3161.6} & \cellcolor[HTML]{C0C0C0} \textbf{3440.3} & \cellcolor[HTML]{C0C0C0} \textbf{3699.4} \\ \hline 
\hline 
\textbf{Case} & \multicolumn{10}{c|}{\textbf{case60api}}\\ \hline 
\textbf{Damage} & \multicolumn{1}{c|}{\textbf{10}}& \multicolumn{1}{c|}{\textbf{20}}& \multicolumn{1}{c|}{\textbf{30}}& \multicolumn{1}{c|}{\textbf{40}}& \multicolumn{1}{c|}{\textbf{50}}& \multicolumn{1}{c|}{\textbf{60}}& \multicolumn{1}{c|}{\textbf{70}}& \multicolumn{1}{c|}{\textbf{80}}& \multicolumn{1}{c|}{\textbf{90}}& \multicolumn{1}{c|}{\textbf{100}}\\ \hline 
\textbf{Power Demand} & \multicolumn{1}{c|}{\textbf{1262.0}}& \multicolumn{1}{c|}{\textbf{2524.0}}& \multicolumn{1}{c|}{\textbf{3645.8}}& \multicolumn{1}{c|}{\textbf{4907.8}}& \multicolumn{1}{c|}{\textbf{6169.8}}& \multicolumn{1}{c|}{\textbf{7431.9}}& \multicolumn{1}{c|}{\textbf{8693.9}}& \multicolumn{1}{c|}{\textbf{9815.7}}& \multicolumn{1}{c|}{\textbf{11077.7}}& \multicolumn{1}{c|}{\textbf{12339.7}}\\ \hline 
\emph{Util}& 1170.9 & 2290.3 & 3261.2 & 3758.9 & 5020.3 & 5221.4 & 4227.1 & 5958.3 & 3971.4 & 4098.4 \\ \hline 
\emph{RAD}& 1218.1 & 2390.0 & 3395.6 & \cellcolor[HTML]{C0C0C0} \textbf{4459.3} & 5689.2 & 6604.1 & \cellcolor[HTML]{C0C0C0} \textbf{7712.6} & 8520.1 & \cellcolor[HTML]{C0C0C0} \textbf{9359.3} & 8983.6 \\ \hline 
\emph{RRR}& \cellcolor[HTML]{C0C0C0} \textbf{1235.1} & \cellcolor[HTML]{C0C0C0} \textbf{2440.4} & \cellcolor[HTML]{C0C0C0} \textbf{3505.0} & \cellcolor[HTML]{C0C0C0} \textbf{4431.9} & \cellcolor[HTML]{C0C0C0} \textbf{5723.9} & \cellcolor[HTML]{C0C0C0} \textbf{6703.6} & \cellcolor[HTML]{C0C0C0} \textbf{7713.4} & \cellcolor[HTML]{C0C0C0} \textbf{8584.8} & \cellcolor[HTML]{C0C0C0} \textbf{9325.3} & \cellcolor[HTML]{C0C0C0} \textbf{10351.6} \\ \hline 
\emph{ROP}& \cellcolor[HTML]{C0C0C0} \textbf{1236.6} & \cellcolor[HTML]{C0C0C0} \textbf{2447.8} & \cellcolor[HTML]{C0C0C0} \textbf{3510.1} & \cellcolor[HTML]{C0C0C0} \textbf{4461.5} & \cellcolor[HTML]{C0C0C0} \textbf{5758.4} & \cellcolor[HTML]{C0C0C0} \textbf{6748.1} & \cellcolor[HTML]{C0C0C0} \textbf{7753.1} & \cellcolor[HTML]{C0C0C0} \textbf{8621.1} & \cellcolor[HTML]{C0C0C0} \textbf{9389.1} & \cellcolor[HTML]{C0C0C0} \textbf{10396.4} \\ \hline 
\hline 
\textbf{Case} & \multicolumn{10}{c|}{\textbf{case118api}}\\ \hline 
\textbf{Damage} & \multicolumn{1}{c|}{\textbf{10}}& \multicolumn{1}{c|}{\textbf{20}}& \multicolumn{1}{c|}{\textbf{30}}& \multicolumn{1}{c|}{\textbf{40}}& \multicolumn{1}{c|}{\textbf{50}}& \multicolumn{1}{c|}{\textbf{60}}& \multicolumn{1}{c|}{\textbf{70}}& \multicolumn{1}{c|}{\textbf{80}}& \multicolumn{1}{c|}{\textbf{90}}& \multicolumn{1}{c|}{\textbf{100}}\\ \hline 
\textbf{Power Demand} & \multicolumn{1}{c|}{\textbf{1307.4}}& \multicolumn{1}{c|}{\textbf{2546.0}}& \multicolumn{1}{c|}{\textbf{3853.3}}& \multicolumn{1}{c|}{\textbf{5091.9}}& \multicolumn{1}{c|}{\textbf{6399.3}}& \multicolumn{1}{c|}{\textbf{7706.7}}& \multicolumn{1}{c|}{\textbf{8945.2}}& \multicolumn{1}{c|}{\textbf{10252.6}}& \multicolumn{1}{c|}{\textbf{11491.2}}& \multicolumn{1}{c|}{\textbf{12798.6}}\\ \hline 
\emph{Util}& \cellcolor[HTML]{C0C0C0} \textbf{1294.6} & 2372.0 & 3480.7 & 4268.0 & 5212.2 & 5702.6 & 6229.5 & 6411.3 & 7196.3 & 7525.4 \\ \hline 
\emph{RAD}& \cellcolor[HTML]{C0C0C0} \textbf{1301.9} & \cellcolor[HTML]{C0C0C0} \textbf{2509.9} & 3673.1 & 4563.7 & \cellcolor[HTML]{C0C0C0} \textbf{6058.6} & \cellcolor[HTML]{C0C0C0} \textbf{7067.4} & 7986.9 & 8792.9 & \cellcolor[HTML]{C0C0C0} \textbf{9823.0} & \cellcolor[HTML]{C0C0C0} \textbf{10599.1} \\ \hline 
\emph{RRR}& \cellcolor[HTML]{C0C0C0} \textbf{1302.3} & \cellcolor[HTML]{C0C0C0} \textbf{2498.6} & \cellcolor[HTML]{C0C0C0} \textbf{3788.4} & \cellcolor[HTML]{C0C0C0} \textbf{4902.9} & \cellcolor[HTML]{C0C0C0} \textbf{6049.8} & \cellcolor[HTML]{C0C0C0} \textbf{7025.6} & \cellcolor[HTML]{C0C0C0} \textbf{8079.3} & \cellcolor[HTML]{C0C0C0} \textbf{8911.1} & \cellcolor[HTML]{C0C0C0} \textbf{9860.8} & \cellcolor[HTML]{C0C0C0} \textbf{10705.2} \\ \hline 
\emph{ROP}& \cellcolor[HTML]{C0C0C0} \textbf{1294.6} & \cellcolor[HTML]{C0C0C0} \textbf{2506.4} & \cellcolor[HTML]{C0C0C0} \textbf{3784.1} & \cellcolor[HTML]{C0C0C0} \textbf{4940.3} & \cellcolor[HTML]{C0C0C0} \textbf{6082.6} & \cellcolor[HTML]{C0C0C0} \textbf{7079.6} & \cellcolor[HTML]{C0C0C0} \textbf{8031.1} & 8426.1 & 8724.8 & 9328.3 \\ \hline 
\hline 
\textbf{Case} & \multicolumn{10}{c|}{\textbf{case240api}}\\ \hline 
\textbf{Damage} & \multicolumn{1}{c|}{\textbf{10}}& \multicolumn{1}{c|}{\textbf{20}}& \multicolumn{1}{c|}{\textbf{30}}& \multicolumn{1}{c|}{\textbf{40}}& \multicolumn{1}{c|}{\textbf{50}}& \multicolumn{1}{c|}{\textbf{60}}& \multicolumn{1}{c|}{\textbf{70}}& \multicolumn{1}{c|}{\textbf{80}}& \multicolumn{1}{c|}{\textbf{90}}& \multicolumn{1}{c|}{\textbf{100}}\\ \hline 
\textbf{Power Demand} & \multicolumn{1}{c|}{\textbf{81410.2}}& \multicolumn{1}{c|}{\textbf{162820.4}}& \multicolumn{1}{c|}{\textbf{242421.5}}& \multicolumn{1}{c|}{\textbf{323831.7}}& \multicolumn{1}{c|}{\textbf{405241.9}}& \multicolumn{1}{c|}{\textbf{486652.1}}& \multicolumn{1}{c|}{\textbf{568062.3}}& \multicolumn{1}{c|}{\textbf{647663.4}}& \multicolumn{1}{c|}{\textbf{729073.6}}& \multicolumn{1}{c|}{\textbf{810483.8}}\\ \hline 
\emph{Util}& \cellcolor[HTML]{C0C0C0} \textbf{81107.0} & 159040.3 & 200671.3 & 257979.8 & 244205.4 & 258124.9 & 118857.2 & 230092.7 & 98234.4 & 58545.0 \\ \hline 
\emph{RAD}& \cellcolor[HTML]{C0C0C0} \textbf{81357.0} & 160091.4 & 235764.4 & \cellcolor[HTML]{C0C0C0} \textbf{315823.7} & 380102.9 & \cellcolor[HTML]{C0C0C0} \textbf{466890.1} & 479083.4 & 590468.9 & 638816.7 & 717060.0 \\ \hline 
\emph{RRR}& \cellcolor[HTML]{C0C0C0} \textbf{81379.2} & \cellcolor[HTML]{C0C0C0} \textbf{162341.7} & \cellcolor[HTML]{C0C0C0} \textbf{239862.2} & \cellcolor[HTML]{C0C0C0} \textbf{316685.5} & \cellcolor[HTML]{C0C0C0} \textbf{395207.7} & \cellcolor[HTML]{C0C0C0} \textbf{471584.6} & \cellcolor[HTML]{C0C0C0} \textbf{534612.6} & \cellcolor[HTML]{C0C0C0} \textbf{609201.5} & \cellcolor[HTML]{C0C0C0} \textbf{675046.1} & \cellcolor[HTML]{C0C0C0} \textbf{739336.7} \\ \hline 
\emph{ROP}& \cellcolor[HTML]{C0C0C0} \textbf{81107.0} & \cellcolor[HTML]{C0C0C0} \textbf{162463.0} & \cellcolor[HTML]{C0C0C0} \textbf{240101.0} & \cellcolor[HTML]{C0C0C0} \textbf{317677.8} & \cellcolor[HTML]{C0C0C0} \textbf{393652.2} & 427799.9 & 471192.5 & 542647.8 & 98234.5 & 58545.0 \\ \hline 
\hline 
\textbf{Case} & \multicolumn{10}{c|}{\textbf{case500gocapi}}\\ \hline 
\textbf{Damage} & \multicolumn{1}{c|}{\textbf{10}}& \multicolumn{1}{c|}{\textbf{20}}& \multicolumn{1}{c|}{\textbf{30}}& \multicolumn{1}{c|}{\textbf{40}}& \multicolumn{1}{c|}{\textbf{50}}& \multicolumn{1}{c|}{\textbf{60}}& \multicolumn{1}{c|}{\textbf{70}}& \multicolumn{1}{c|}{\textbf{80}}& \multicolumn{1}{c|}{\textbf{90}}& \multicolumn{1}{c|}{\textbf{100}}\\ \hline 
\textbf{Power Demand} & \multicolumn{1}{c|}{\textbf{20146.1}}& \multicolumn{1}{c|}{\textbf{40292.2}}& \multicolumn{1}{c|}{\textbf{59886.4}}& \multicolumn{1}{c|}{\textbf{80308.4}}& \multicolumn{1}{c|}{\textbf{100178.6}}& \multicolumn{1}{c|}{\textbf{120600.6}}& \multicolumn{1}{c|}{\textbf{140194.8}}& \multicolumn{1}{c|}{\textbf{160616.9}}& \multicolumn{1}{c|}{\textbf{181038.9}}& \multicolumn{1}{c|}{\textbf{200909.1}}\\ \hline 
\emph{Util}& 18461.1 & 34984.5 & 46534.6 & 53635.6 & 59729.2 & 57430.5 & 59384.8 & 64958.6 & 56922.7 & 58043.0 \\ \hline 
\emph{RAD}& 19335.6 & 37336.2 & 55795.0 & 72550.5 & 86519.0 & 97541.1 & 99325.3 & 100699.1 & 78213.2 & 86193.6 \\ \hline 
\emph{RRR}& \cellcolor[HTML]{C0C0C0} \textbf{19874.6} & \cellcolor[HTML]{C0C0C0} \textbf{39287.9} & \cellcolor[HTML]{C0C0C0} \textbf{56940.9} & \cellcolor[HTML]{C0C0C0} \textbf{74415.7} & \cellcolor[HTML]{C0C0C0} \textbf{89540.5} & \cellcolor[HTML]{C0C0C0} \textbf{104426.3} & \cellcolor[HTML]{C0C0C0} \textbf{117474.0} & \cellcolor[HTML]{C0C0C0} \textbf{131775.4} & \cellcolor[HTML]{C0C0C0} \textbf{143194.1} & \cellcolor[HTML]{C0C0C0} \textbf{154440.2} \\ \hline 
\emph{ROP}& \cellcolor[HTML]{C0C0C0} \textbf{19845.3} & \cellcolor[HTML]{C0C0C0} \textbf{39399.1} & 55209.9 & 53635.6 & 59729.2 & 57430.5 & 59384.8 & 64958.6 & 56922.7 & 58043.0 \\ \hline 
\end{tabular} \\  
 
\label{tab:ENS_10h_data}
\end{table*}

\begin{table*}[t]
\centering
\caption{Total Power Served [p.u.] with 5 minute algorithm time limit.}
\begin{tabular}{|l||r|r|r|r|r|r|r|r|r|r|}
\hline 
\textbf{Case} & \multicolumn{10}{c|}{\textbf{case24api}}\\ \hline 
\textbf{Damage} & \multicolumn{1}{c|}{\textbf{10}}& \multicolumn{1}{c|}{\textbf{20}}& \multicolumn{1}{c|}{\textbf{30}}& \multicolumn{1}{c|}{\textbf{40}}& \multicolumn{1}{c|}{\textbf{50}}& \multicolumn{1}{c|}{\textbf{60}}& \multicolumn{1}{c|}{\textbf{70}}& \multicolumn{1}{c|}{\textbf{80}}& \multicolumn{1}{c|}{\textbf{90}}& \multicolumn{1}{c|}{\textbf{100}}\\ \hline 
\textbf{Power Demand} & \multicolumn{1}{c|}{\textbf{218.8}}& \multicolumn{1}{c|}{\textbf{437.6}}& \multicolumn{1}{c|}{\textbf{601.7}}& \multicolumn{1}{c|}{\textbf{820.6}}& \multicolumn{1}{c|}{\textbf{1039.4}}& \multicolumn{1}{c|}{\textbf{1258.2}}& \multicolumn{1}{c|}{\textbf{1477.0}}& \multicolumn{1}{c|}{\textbf{1641.1}}& \multicolumn{1}{c|}{\textbf{1859.9}}& \multicolumn{1}{c|}{\textbf{2078.8}}\\ \hline 
\emph{Util}& 216.5 & \cellcolor[HTML]{C0C0C0} \textbf{434.5} & 590.7 & 771.0 & 918.0 & 1095.0 & 1319.1 & 1353.6 & 1410.9 & 1521.8 \\ \hline 
\emph{RAD}& \cellcolor[HTML]{C0C0C0} \textbf{218.8} & \cellcolor[HTML]{C0C0C0} \textbf{435.8} & \cellcolor[HTML]{C0C0C0} \textbf{597.2} & \cellcolor[HTML]{C0C0C0} \textbf{801.1} & \cellcolor[HTML]{C0C0C0} \textbf{998.0} & \cellcolor[HTML]{C0C0C0} \textbf{1192.5} & \cellcolor[HTML]{C0C0C0} \textbf{1425.6} & 1457.6 & 1578.5 & 1834.0 \\ \hline 
\emph{RRR}& \cellcolor[HTML]{C0C0C0} \textbf{218.8} & \cellcolor[HTML]{C0C0C0} \textbf{436.0} & \cellcolor[HTML]{C0C0C0} \textbf{596.8} & \cellcolor[HTML]{C0C0C0} \textbf{801.1} & \cellcolor[HTML]{C0C0C0} \textbf{1003.0} & \cellcolor[HTML]{C0C0C0} \textbf{1194.6} & \cellcolor[HTML]{C0C0C0} \textbf{1426.1} & \cellcolor[HTML]{C0C0C0} \textbf{1524.8} & \cellcolor[HTML]{C0C0C0} \textbf{1668.6} & \cellcolor[HTML]{C0C0C0} \textbf{1860.2} \\ \hline 
\emph{ROP}& 216.5 & \cellcolor[HTML]{C0C0C0} \textbf{434.5} & \cellcolor[HTML]{C0C0C0} \textbf{596.0} & \cellcolor[HTML]{C0C0C0} \textbf{800.4} & \cellcolor[HTML]{C0C0C0} \textbf{1000.7} & \cellcolor[HTML]{C0C0C0} \textbf{1200.4} & \cellcolor[HTML]{C0C0C0} \textbf{1424.8} & \cellcolor[HTML]{C0C0C0} \textbf{1530.5} & \cellcolor[HTML]{C0C0C0} \textbf{1679.4} & \cellcolor[HTML]{C0C0C0} \textbf{1864.4} \\ \hline 
\hline 
\textbf{Case} & \multicolumn{10}{c|}{\textbf{case39api}}\\ \hline 
\textbf{Damage} & \multicolumn{1}{c|}{\textbf{10}}& \multicolumn{1}{c|}{\textbf{20}}& \multicolumn{1}{c|}{\textbf{30}}& \multicolumn{1}{c|}{\textbf{40}}& \multicolumn{1}{c|}{\textbf{50}}& \multicolumn{1}{c|}{\textbf{60}}& \multicolumn{1}{c|}{\textbf{70}}& \multicolumn{1}{c|}{\textbf{80}}& \multicolumn{1}{c|}{\textbf{90}}& \multicolumn{1}{c|}{\textbf{100}}\\ \hline 
\textbf{Power Demand} & \multicolumn{1}{c|}{\textbf{505.3}}& \multicolumn{1}{c|}{\textbf{909.6}}& \multicolumn{1}{c|}{\textbf{1415.0}}& \multicolumn{1}{c|}{\textbf{1819.2}}& \multicolumn{1}{c|}{\textbf{2324.6}}& \multicolumn{1}{c|}{\textbf{2829.9}}& \multicolumn{1}{c|}{\textbf{3234.2}}& \multicolumn{1}{c|}{\textbf{3739.6}}& \multicolumn{1}{c|}{\textbf{4143.8}}& \multicolumn{1}{c|}{\textbf{4649.2}}\\ \hline 
\emph{Util}& 451.5 & 843.3 & 1193.4 & 1254.6 & 1463.5 & 1612.4 & 1731.6 & 1644.9 & 1810.7 & 1633.1 \\ \hline 
\emph{RAD}& 484.0 & \cellcolor[HTML]{C0C0C0} \textbf{877.3} & \cellcolor[HTML]{C0C0C0} \textbf{1332.6} & 1608.0 & 1726.0 & 2481.8 & 2570.3 & 3128.8 & 3211.9 & 3647.1 \\ \hline 
\emph{RRR}& \cellcolor[HTML]{C0C0C0} \textbf{492.5} & \cellcolor[HTML]{C0C0C0} \textbf{876.3} & \cellcolor[HTML]{C0C0C0} \textbf{1333.5} & \cellcolor[HTML]{C0C0C0} \textbf{1669.4} & \cellcolor[HTML]{C0C0C0} \textbf{1994.5} & \cellcolor[HTML]{C0C0C0} \textbf{2490.7} & \cellcolor[HTML]{C0C0C0} \textbf{2694.5} & \cellcolor[HTML]{C0C0C0} \textbf{3133.1} & \cellcolor[HTML]{C0C0C0} \textbf{3414.4} & \cellcolor[HTML]{C0C0C0} \textbf{3667.7} \\ \hline 
\emph{ROP}& \cellcolor[HTML]{C0C0C0} \textbf{488.4} & \cellcolor[HTML]{C0C0C0} \textbf{876.7} & \cellcolor[HTML]{C0C0C0} \textbf{1339.1} & \cellcolor[HTML]{C0C0C0} \textbf{1668.6} & \cellcolor[HTML]{C0C0C0} \textbf{2011.2} & \cellcolor[HTML]{C0C0C0} \textbf{2511.3} & \cellcolor[HTML]{C0C0C0} \textbf{2721.2} & \cellcolor[HTML]{C0C0C0} \textbf{3161.6} & \cellcolor[HTML]{C0C0C0} \textbf{3440.3} & \cellcolor[HTML]{C0C0C0} \textbf{3699.2} \\ \hline 
\hline 
\textbf{Case} & \multicolumn{10}{c|}{\textbf{case60api}}\\ \hline 
\textbf{Damage} & \multicolumn{1}{c|}{\textbf{10}}& \multicolumn{1}{c|}{\textbf{20}}& \multicolumn{1}{c|}{\textbf{30}}& \multicolumn{1}{c|}{\textbf{40}}& \multicolumn{1}{c|}{\textbf{50}}& \multicolumn{1}{c|}{\textbf{60}}& \multicolumn{1}{c|}{\textbf{70}}& \multicolumn{1}{c|}{\textbf{80}}& \multicolumn{1}{c|}{\textbf{90}}& \multicolumn{1}{c|}{\textbf{100}}\\ \hline 
\textbf{Power Demand} & \multicolumn{1}{c|}{\textbf{1262.0}}& \multicolumn{1}{c|}{\textbf{2524.0}}& \multicolumn{1}{c|}{\textbf{3645.8}}& \multicolumn{1}{c|}{\textbf{4907.8}}& \multicolumn{1}{c|}{\textbf{6169.8}}& \multicolumn{1}{c|}{\textbf{7431.9}}& \multicolumn{1}{c|}{\textbf{8693.9}}& \multicolumn{1}{c|}{\textbf{9815.7}}& \multicolumn{1}{c|}{\textbf{11077.7}}& \multicolumn{1}{c|}{\textbf{12339.7}}\\ \hline 
\emph{Util}& 1170.9 & 2290.3 & 3261.2 & 3758.9 & 5020.3 & 5221.4 & 4227.1 & 5958.3 & 3971.4 & 4098.4 \\ \hline 
\emph{RAD}& 1218.1 & 2389.5 & 3394.7 & \cellcolor[HTML]{C0C0C0} \textbf{4447.8} & 5684.8 & 6574.8 & \cellcolor[HTML]{C0C0C0} \textbf{7692.1} & 8474.6 & \cellcolor[HTML]{C0C0C0} \textbf{9328.2} & 8897.2 \\ \hline 
\emph{RRR}& \cellcolor[HTML]{C0C0C0} \textbf{1235.1} & \cellcolor[HTML]{C0C0C0} \textbf{2440.4} & \cellcolor[HTML]{C0C0C0} \textbf{3505.0} & \cellcolor[HTML]{C0C0C0} \textbf{4431.9} & \cellcolor[HTML]{C0C0C0} \textbf{5723.9} & \cellcolor[HTML]{C0C0C0} \textbf{6703.6} & \cellcolor[HTML]{C0C0C0} \textbf{7713.4} & \cellcolor[HTML]{C0C0C0} \textbf{8584.8} & \cellcolor[HTML]{C0C0C0} \textbf{9325.3} & \cellcolor[HTML]{C0C0C0} \textbf{10351.6} \\ \hline 
\emph{ROP}& \cellcolor[HTML]{C0C0C0} \textbf{1236.6} & \cellcolor[HTML]{C0C0C0} \textbf{2447.8} & \cellcolor[HTML]{C0C0C0} \textbf{3510.1} & \cellcolor[HTML]{C0C0C0} \textbf{4461.5} & \cellcolor[HTML]{C0C0C0} \textbf{5758.4} & \cellcolor[HTML]{C0C0C0} \textbf{6748.1} & \cellcolor[HTML]{C0C0C0} \textbf{7753.1} & \cellcolor[HTML]{C0C0C0} \textbf{8621.1} & 9185.3 & 9614.0 \\ \hline 
\hline 
\textbf{Case} & \multicolumn{10}{c|}{\textbf{case118api}}\\ \hline 
\textbf{Damage} & \multicolumn{1}{c|}{\textbf{10}}& \multicolumn{1}{c|}{\textbf{20}}& \multicolumn{1}{c|}{\textbf{30}}& \multicolumn{1}{c|}{\textbf{40}}& \multicolumn{1}{c|}{\textbf{50}}& \multicolumn{1}{c|}{\textbf{60}}& \multicolumn{1}{c|}{\textbf{70}}& \multicolumn{1}{c|}{\textbf{80}}& \multicolumn{1}{c|}{\textbf{90}}& \multicolumn{1}{c|}{\textbf{100}}\\ \hline 
\textbf{Power Demand} & \multicolumn{1}{c|}{\textbf{1307.4}}& \multicolumn{1}{c|}{\textbf{2546.0}}& \multicolumn{1}{c|}{\textbf{3853.3}}& \multicolumn{1}{c|}{\textbf{5091.9}}& \multicolumn{1}{c|}{\textbf{6399.3}}& \multicolumn{1}{c|}{\textbf{7706.7}}& \multicolumn{1}{c|}{\textbf{8945.2}}& \multicolumn{1}{c|}{\textbf{10252.6}}& \multicolumn{1}{c|}{\textbf{11491.2}}& \multicolumn{1}{c|}{\textbf{12798.6}}\\ \hline 
\emph{Util}& \cellcolor[HTML]{C0C0C0} \textbf{1294.6} & 2372.0 & 3480.7 & 4268.0 & 5212.2 & 5702.6 & 6229.5 & 6411.3 & 7196.3 & 7525.4 \\ \hline 
\emph{RAD}& \cellcolor[HTML]{C0C0C0} \textbf{1299.9} & \cellcolor[HTML]{C0C0C0} \textbf{2509.7} & 3631.3 & 4537.6 & 5520.2 & 6378.7 & 7300.6 & 7469.5 & 8102.6 & 8483.4 \\ \hline 
\emph{RRR}& \cellcolor[HTML]{C0C0C0} \textbf{1302.3} & \cellcolor[HTML]{C0C0C0} \textbf{2498.6} & \cellcolor[HTML]{C0C0C0} \textbf{3788.4} & \cellcolor[HTML]{C0C0C0} \textbf{4902.9} & \cellcolor[HTML]{C0C0C0} \textbf{6049.8} & \cellcolor[HTML]{C0C0C0} \textbf{7025.6} & \cellcolor[HTML]{C0C0C0} \textbf{8079.3} & \cellcolor[HTML]{C0C0C0} \textbf{8911.1} & \cellcolor[HTML]{C0C0C0} \textbf{9860.8} & \cellcolor[HTML]{C0C0C0} \textbf{10705.2} \\ \hline 
\emph{ROP}& \cellcolor[HTML]{C0C0C0} \textbf{1294.6} & \cellcolor[HTML]{C0C0C0} \textbf{2506.4} & \cellcolor[HTML]{C0C0C0} \textbf{3784.1} & 4795.0 & 5827.1 & 5702.6 & 6229.5 & 6411.3 & 7196.3 & 7525.4 \\ \hline 
\hline 
\textbf{Case} & \multicolumn{10}{c|}{\textbf{case240api}}\\ \hline 
\textbf{Damage} & \multicolumn{1}{c|}{\textbf{10}}& \multicolumn{1}{c|}{\textbf{20}}& \multicolumn{1}{c|}{\textbf{30}}& \multicolumn{1}{c|}{\textbf{40}}& \multicolumn{1}{c|}{\textbf{50}}& \multicolumn{1}{c|}{\textbf{60}}& \multicolumn{1}{c|}{\textbf{70}}& \multicolumn{1}{c|}{\textbf{80}}& \multicolumn{1}{c|}{\textbf{90}}& \multicolumn{1}{c|}{\textbf{100}}\\ \hline 
\textbf{Power Demand} & \multicolumn{1}{c|}{\textbf{81410.2}}& \multicolumn{1}{c|}{\textbf{162820.4}}& \multicolumn{1}{c|}{\textbf{242421.5}}& \multicolumn{1}{c|}{\textbf{323831.7}}& \multicolumn{1}{c|}{\textbf{405241.9}}& \multicolumn{1}{c|}{\textbf{486652.1}}& \multicolumn{1}{c|}{\textbf{568062.3}}& \multicolumn{1}{c|}{\textbf{647663.4}}& \multicolumn{1}{c|}{\textbf{729073.6}}& \multicolumn{1}{c|}{\textbf{810483.8}}\\ \hline 
\emph{Util}& \cellcolor[HTML]{C0C0C0} \textbf{81107.0} & 159040.3 & 200671.3 & 257979.8 & 244205.4 & 258124.9 & 118857.2 & 230092.7 & 98234.4 & 58545.0 \\ \hline 
\emph{RAD}& \cellcolor[HTML]{C0C0C0} \textbf{81180.6} & 159586.0 & 213379.3 & 264473.5 & 260538.3 & 266645.0 & 133016.9 & 240215.5 & 113153.4 & 77605.7 \\ \hline 
\emph{RRR}& \cellcolor[HTML]{C0C0C0} \textbf{81379.2} & \cellcolor[HTML]{C0C0C0} \textbf{162341.7} & \cellcolor[HTML]{C0C0C0} \textbf{239862.2} & \cellcolor[HTML]{C0C0C0} \textbf{316685.5} & \cellcolor[HTML]{C0C0C0} \textbf{395207.7} & \cellcolor[HTML]{C0C0C0} \textbf{471584.6} & \cellcolor[HTML]{C0C0C0} \textbf{534612.6} & \cellcolor[HTML]{C0C0C0} \textbf{609201.5} & \cellcolor[HTML]{C0C0C0} \textbf{675046.1} & \cellcolor[HTML]{C0C0C0} \textbf{739336.7} \\ \hline 
\emph{ROP}& \cellcolor[HTML]{C0C0C0} \textbf{81107.0} & \cellcolor[HTML]{C0C0C0} \textbf{162463.0} & 200671.3 & 257979.8 & 244205.4 & 258124.9 & 118857.2 & 230092.7 & 98234.5 & 58545.0 \\ \hline 
\hline 
\textbf{Case} & \multicolumn{10}{c|}{\textbf{case500gocapi}}\\ \hline 
\textbf{Damage} & \multicolumn{1}{c|}{\textbf{10}}& \multicolumn{1}{c|}{\textbf{20}}& \multicolumn{1}{c|}{\textbf{30}}& \multicolumn{1}{c|}{\textbf{40}}& \multicolumn{1}{c|}{\textbf{50}}& \multicolumn{1}{c|}{\textbf{60}}& \multicolumn{1}{c|}{\textbf{70}}& \multicolumn{1}{c|}{\textbf{80}}& \multicolumn{1}{c|}{\textbf{90}}& \multicolumn{1}{c|}{\textbf{100}}\\ \hline 
\textbf{Power Demand} & \multicolumn{1}{c|}{\textbf{20146.1}}& \multicolumn{1}{c|}{\textbf{40292.2}}& \multicolumn{1}{c|}{\textbf{59886.4}}& \multicolumn{1}{c|}{\textbf{80308.4}}& \multicolumn{1}{c|}{\textbf{100178.6}}& \multicolumn{1}{c|}{\textbf{120600.6}}& \multicolumn{1}{c|}{\textbf{140194.8}}& \multicolumn{1}{c|}{\textbf{160616.9}}& \multicolumn{1}{c|}{\textbf{181038.9}}& \multicolumn{1}{c|}{\textbf{200909.1}}\\ \hline 
\emph{Util}& 18461.1 & 34984.5 & 46534.6 & 53635.6 & 59729.2 & 57430.5 & 59384.8 & 64958.6 & 56922.7 & 58043.0 \\ \hline 
\emph{RAD}& 18794.1 & 35319.8 & 46758.5 & 54069.2 & 60220.6 & 58022.7 & 59911.4 & 65510.9 & 57519.8 & 58642.4 \\ \hline 
\emph{RRR}& \cellcolor[HTML]{C0C0C0} \textbf{19874.6} & \cellcolor[HTML]{C0C0C0} \textbf{39287.9} & \cellcolor[HTML]{C0C0C0} \textbf{56940.9} & \cellcolor[HTML]{C0C0C0} \textbf{74415.7} & \cellcolor[HTML]{C0C0C0} \textbf{89540.5} & \cellcolor[HTML]{C0C0C0} \textbf{104426.3} & \cellcolor[HTML]{C0C0C0} \textbf{117474.0} & \cellcolor[HTML]{C0C0C0} \textbf{131548.1} & \cellcolor[HTML]{C0C0C0} \textbf{142571.5} & \cellcolor[HTML]{C0C0C0} \textbf{154436.1} \\ \hline 
\emph{ROP}& 18461.1 & 34984.5 & 46534.6 & 53635.6 & 59729.2 & 57430.5 & 59384.8 & 64958.6 & 56922.7 & 58043.0 \\ \hline 
\end{tabular} \\  
\label{tab:ENS_5m_data}
\end{table*}

\section{Randomized Adaptive Decomposition}\label{app:RAD}
In this appendix, we summarize the RAD algorithm. Further explanation can be found in \cite{coffrin2012last}.
Algorithm \ref{alg:RAD} shows the RAD algorithm. 
First, an initial restoration sequence $\mathcal{R}$ is found for a power network $\mathcal{N}$ using the Utilization Heuristic.
This restoration sequence is partitioned into subsets of random size, with between $\underline{\boldsymbol{s}}$ and $\overline{\boldsymbol{s}}$ restoration periods in each partition. 
For each partition, an $ROP$ problem is solved with the goal of reordering the restorations to improve the power served. 
If ROP finds a solution that improves the power served, the newly optimized restoration order of the partition $\hat{\mathcal{S}}_n$ is reinserted into the restoration order $\cal R$, and the network is re-partitioned 
again.  

\newpage

The RAD algorithm requires parameters to decide the size of the random partitions, which are adapted across iterations to continue to improve the solutions, as well as parameters to decide how to allocate the allotted computation time among subproblems. 
Here, we initially set the minimum and maximum partition size to $\underline{s}=2$ and $\overline{s}=5$, while the ROP sub-problem time limit is initialized to 1\% of the total time limit.
If 80\% of the partitions fail to improve the solution and more than 80\% of the sub-problem terminate without solving the problem to optimality, then the sub-problem time limit is doubled.
If 80\% of the partitions fail to improve the solution and less than 80\% of the problems are interrupted by the time limit, then the new maximum partition size is increased by 10\% up to a maximum of $|\bcancel{\cal L}|/2$.
The algorithm terminates when it reaches the time limit or if it finished 100 iterations without improving the solution. 

\begin{algorithm}[t]
    \caption{Randomized Adaptive Decomposition (RAD)}
    \begin{algorithmic}[1]
        \renewcommand{\algorithmicrequire}{\textbf{Input:}}
        \renewcommand{\algorithmicensure}{\textbf{Output:}}
        \REQUIRE $\mathcal{N}, \bcancel{\mathcal{L}}$
        \ENSURE  $\mathcal{R}$\\ 
        \STATE $\mathcal{R} \leftarrow UTIL(\mathcal{N}, \bcancel{\mathcal{L}})$ \\
        \WHILE{\NOT $stoppingCriteria$}
            \STATE $\left[ \mathcal{S}_1,...,\mathcal{S}_l\right] \leftarrow RandomPartition(\mathcal{R},[\underline{\boldsymbol{s}}...\overline{\boldsymbol{s}}])$ \\
                \FOR {$\mathcal{S}_n \in \left[ \mathcal{S}_1,...,\mathcal{S}_l\right]$}
                    \STATE $\bcancel{\mathcal{L}}_n \leftarrow \bcancel{\mathcal{L}} \in \mathcal{S}_n$
                    \STATE $\hat{\mathcal{S}}_n \leftarrow$ $ROP(\mathcal{N}, \bcancel{\mathcal{L}}_n, |\mathcal{S}_n|)$
                    \IF {\NOT $(ROP \text{ failure})$ \AND $\hat{\mathcal{S}}_n > {\mathcal{S}}_n$}
                            \STATE $\mathcal{S}_n \leftarrow \hat{\mathcal{S}}_n$
                    \ENDIF
                \ENDFOR
        \ENDWHILE
        \RETURN $\mathcal{R}$ 
    \end{algorithmic} \label{alg:RAD}
\end{algorithm}

\end{document}